\documentclass[12pt,preprint]{aastex}

\shorttitle{BXS Survey - III}
\shortauthors{Ajello et al. 2007b}

\begin{document}



\title{BAT X-ray Survey - III: X-ray Spectra and Statistical 
Properties.}


\author{M. Ajello\altaffilmark{1}, A. Rau\altaffilmark{2},
J. Greiner\altaffilmark{1},
G. Kanbach\altaffilmark{1}, M. Salvato\altaffilmark{2},
A. W. Strong\altaffilmark{1}, S. D. Barthelmy\altaffilmark{3}, 
N. Gehrels\altaffilmark{3}, C. B. Markwardt\altaffilmark{3} 
and J. Tueller\altaffilmark{3}}
\altaffiltext{1}{Max-Planck Institut f\"ur Extraterrestrische Physik, Postfach 1312, 
85741, Garching, Germany}
\altaffiltext{2}{Caltech Optical Observatories, MS 105-24, California, Institute of 
Technology, Pasadena, CA 91125, USA}
\altaffiltext{3}{Astroparticle Physics Laboratory, Mail Code 661, NASA Goddard SpaceFlightCenter, Greenbelt, MD 20771, USA }

\email{majello@mpe.mpg.de}

%
%
%
%
%
%
\begin{abstract}
In this concluding part
of the series of three papers dedicated to the
{\em Swift}/BAT hard X-ray survey (BXS), we focus on 
the X-ray spectral analysis and statistical properties 
of the source sample.
Using a dedicated method to extract  time-averaged spectra of BAT sources
we show that Galactic sources have, generally, softer spectra than
extragalactic objects and that  Seyfert 2 galaxies are harder than Seyfert 1s.
The averaged spectrum of all Seyfert galaxies is consistent
with a power-law with photon index of 2.00$\pm0.07$.
The cumulative flux-number relation 
for the extragalactic sources in the 14-170\,keV band
is best described by a power-law with a slope 
$\alpha=1.55\pm0.20$ and
a normalization of 9.6$\pm1.9 \times 10^{-3}$
AGN deg$^{-2}$ (or 396$\pm 80$ AGN all-sky)  
above a  flux level of 2$\times 10^{-11}$erg cm$^{-2}$ s$^{-1}$ 
($\sim$0.85 mCrab).
The integration of the 
cumulative flux per unit area indicates that BAT resolves
1-2\% of the X-ray background emission in the 14-170\,keV band. 
A sub-sample of 24 extragalactic sources above the 4.5\,$\sigma$ detection limit 
is used to study the statistical 
properties of AGN. 
This  sample  comprises local Seyfert 
galaxies (z=0.026, median value)
and  $\sim$10\% blazars.
We find that  55\% of the Seyfert galaxies are absorbed
by column densities of N$_H>10^{22}$\,H-atoms cm$^{-2}$, but that none is
a bona fide  Compton-thick.
This study shows the capabilities of BAT to probe the hard X-ray sky to
the mCrab level.
\end{abstract}


\section{Introduction}
There is a general consensus that the cosmic X-ray
background (CXB), discovered more than 40 years ago \citep{giacconi62},
is produced by integrated emission of 
Active Galactic Nuclei (AGN).
Population synthesis models have successfully shown, 
in the context of the AGN unified theory \citep{antonucci93},
that  AGN  with various level of obscuration and at different redshifts 
account for  80--100\% of the CXB below 4\,keV
\citep{comastri95,gilli01,treister05}.
Notwithstanding all the advances in the field a major question remains,
do Compton-thick sources exist in the numbers that seem to be required
by population synthesis models \citep[e.g.][]{comastri95,gilli01}
to reproduce the shape of the CXB emission?
Indication of the existence of such population comes from the 
analysis of the CXB fraction which is   resolved-into-sources;
\cite{worsley05} find that such fraction decreases with energy
and that 
the unresolved component is consistent as 
being the emission of a yet undetected population of Compton-thick AGN.
In synthesis, much evidence points towards the existence of Compton-thick
AGN while only a handful of them are known and studied.

The $>$10\,keV energy range is the most appropriate band for studying
and selecting an unbiased (with respect to absorption) sample of AGN.
This band is also the optimum band for the detection of Compton-thick objects.
These elusive objects could have been missed
because of the difficulties of performing
sensitive imaging of the hard X-ray sky.
The Burst Alert Telescope (BAT) \citep{barthelmy05}, on board the 
Swift mission \citep{gehrels04}
 represents 
a major improvements in sensitivity for X-ray imaging of the hard X-ray sky.
We refer to \cite{ajello07a} for details about the  BXS survey.
\\\\
We applied an innovative image reconstruction algorithm
to 8 months of survey BAT data; our survey 
covers $\sim$7000\,deg$^{2}$
reaching a limiting  sensitivity of $<0.9$ mCrab.
This makes it one of the most sensitive survey ever performed in the hard
X-ray domain. We detected 49 hard X-ray sources of which 37 were
previously unknown as hard X-ray emitters. Correlation with X-ray
catalogs allowed us to identify $~$15 sources, while {\em Swift}/XRT
pointed observations provided identification for another 15 objects.
Furthermore, we optically identified
three new  extragalactic sources \citep{rau07}.
Here, we investigate the spectral and statistical properties of all
complete source sample.
\\\\
The paper is organized as follows. In section \ref{sec:spec}
we present the X-ray spectral analysis of the BAT sources.
The details of the dedicated spectral extraction method are presented in
the Appendix \ref{app:spec}.
We  use the source spectra to build an X-ray color-color
plot which is used to understand the mean properties of the source populations.
In section \ref{sec:extra}, we apply the V/V$_{MAX}$ method to test
the completeness of the extragalactic sample which is then used
to derive the number-flux relation. 
The section
ends with a discussion about the statistical properties of the 
extragalactic sample.
Finally, we discuss the BAT results in section \ref{sec:disc}. 
Throughout this work we use  $H_0$ = 70\,km s$^{-1}$ Mpc$^{-1}$
(h$_{70}$ =1), $k=0$, $\Omega_{matter}$=0.3 and $\Lambda_0$=0.7 
and the luminosities are given in erg s$^{-1}$ h$^{-1}$.

%
%
\section{Spectral analysis}\label{sec:spec}
We have developed a dedicated spectral extraction method which 
allows to derive the time-averaged spectrum of all sources.
The reader interested in the method is referred
to the Appendix \ref{app:extr} for details.
Using this method, we derived 
for all our source candidates a 6 channels energy spectrum
in the 14-195\,keV. The energy channels used in this analysis
are (in keV): 14--22, 22--30, 30--47,  47--71, 71-121, 121--195.
The energy bins were optimally chosen to produce similar error bars
(in the different energy bins) for sources with power-law spectra.
\\
We found that 
21 sources had at least soft X-ray observations by {\em Swift}/XRT or ASCA.  
For these sources, we jointly fit XRT/ASCA and BAT data.
When fitting a source spectrum, we have preferred the simplest model
yielding a good description of the data. The normalization
of the ASCA spectra was allowed to vary (with respect to the BAT ones) 
to cope with the different epochs of the observations. This was not 
required when fitting XRT and BAT data.
In general, the BAT spectrum
of Galactic sources is well fitted by a thermal bremsstrahlung model.
Instead,  AGN are usually better described by a single power-law model.
However, when $<$10\,keV data were available the fit required additional 
components (i.e. black body component for soft excess and/or gaussian model
for the iron line).
The detailed analysis is reported in Appendix \ref{app:spec} while
the spectral parameters are summarized in Table \ref{tab:spec}.

%
%

\begin{figure}[h!]
\epsscale{1.}
\plotone{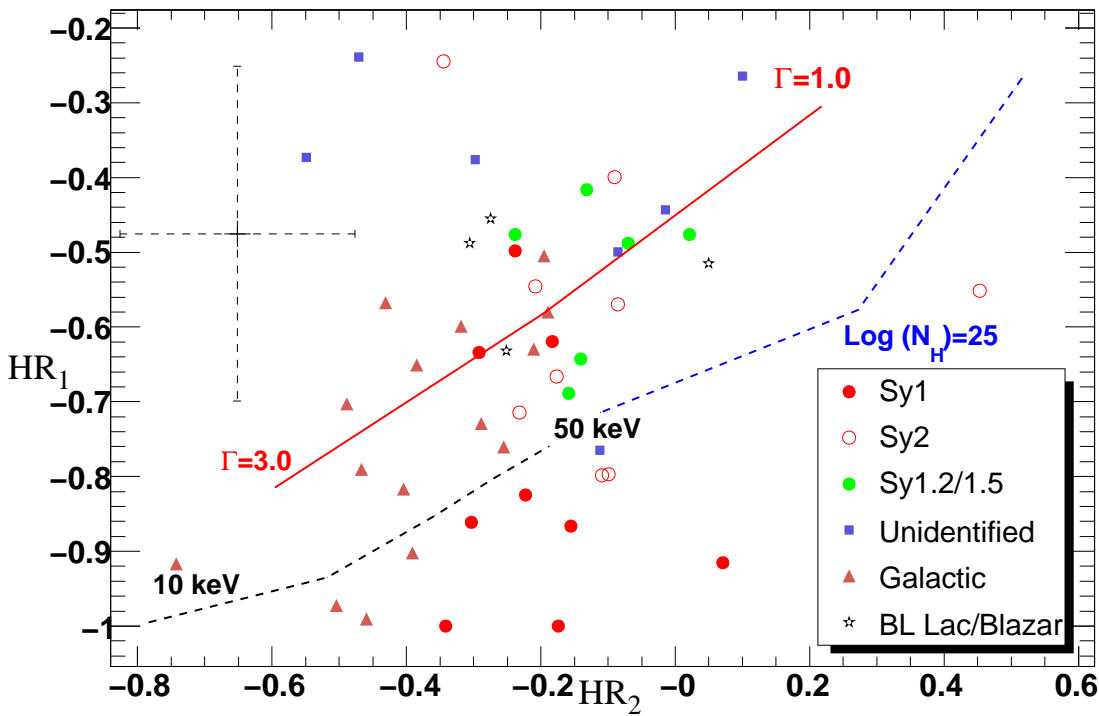}
	\caption{Plot of HR1 and HR2 hardness ratios.
The solid line is the locus for sources
with unabsorbed power law spectra with photon indices from 1.0 to 3.0
while the long dashed line shows the location of Compton-thick AGN with
the same range of photon indices.
The dashed line shows the location
of objects with a thermal bremsstrahlung spectrum with temperatures in the 
range 5--50\,keV. In the upper left corner the typical $\pm1\sigma$
error for a 5\,$\sigma$ source is shown. 
}
	\label{fig:HR}
\end{figure}

%
%
The properties of the source sample can be studied using  hardness ratios.
We have, thus, defined HR$_1$ and HR$_2$ as:
\begin{equation}
HR_1=\frac{medium-hard}{medium+hard}, \ \ \ \ 
HR_2=\frac{soft-medium} {soft+medium},
\end{equation}

where the {\it soft, medium} and {\it hard} bands are respectively (in keV):
14--30, 30--71, 71--195.
The hardness ratios, shown in Fig.\ref{fig:HR}, are normalized to the range -1 and +1.
Different symbols indicate different source classes.
We also indicate  the loci
occupied by sources with a power law index in the range 1.0-3.5,
or a bremsstrahlung spectrum with a temperature
of 5--50\,keV. 
A few things can be derived by the study of the hardness ratios.
Galactic sources, which are usually characterized by soft X-ray spectra,
have $HR_2$ values $<-0.3$  and $HR_1<-0.5$ which is the typical region
for sources with steep photon index. Indeed, the five Cataclysmic Variables (CVs)
present in the sample are all well fit by a relativistic
bremsstrahlung model with  a mean plasma temperature of 23\,keV.

Similarly we note that Sy2 galaxies seem to have (given the large
uncertainties)  harder X-ray spectra than Sy1s (larger values of $HR_1$).
The fact that type-2 AGN have systematically harder spectra than type-1
could be an evidence of the intrinsic difference between these two
classes of objects. In order to study this in more detail we performed
a stacked spectral analysis\footnotemark{}
\footnotetext{All stacked spectral 
analysis are performed doing a weighted average
of the spectra.} 
grouping the Seyfert galaxies detected by BAT into
three classes: Seyfert 1, Seyfert 2 and intermediate Seyferts.
The results, which are summarized in Table \ref{tab:class},
show that the mean photon index of Seyfert-1 and Seyfert-2 are different
at more than 2\,$\sigma$ level. The same trend was previously noted in Seyfert
galaxies detected by {\it OSSE} \citep{zdziarski00} and by INTEGRAL 
\citep{beckmann06}.
\cite{zdziarski00} find that the difference in 
spectral index could be due to the different viewing angle between
Seyfert 1s and 2s. Indeed, the strength of Compton reflection
decreases with the increasing viewing angle. Since the spectrum from
Compton reflection peaks at 30\,keV followed by a steep decline, the larger
the reflection component, the softer the spectrum. We tested this scenario
using for Seyfert 1s (Seyfert 2s are successfully fitted by a simple power-law)  the PEXRAV \citep{magdziarz95} model in XSPEC. 
Indeed we get a good fit ($\chi^2=1.2/3$) 
with a (minimum) reflection strength R$>1.1$ (upper limit is unconstrained
by the fit) which is in good agreement
with what found by  \cite{zdziarski00} and \cite{deluit03}. 
Thus, the BAT data seem to confirm
the larger reflection component present in Seyfert 1 galaxies (with respect to
Seyfert 2s) in agreement with the AGN unified model.
Even though, the reflection component improves the fit, it does not affect
the photon index of Seyfert 1s which remains  2.30$\pm0.12$.

We note, however, that most of the Seyfert 1s (6 out of 9) have
a  low value of HR$_1$ denoting a steep spectrum. We thus tried to fit
the stacked spectrum with a cutoff power-law model of the form 
${\textup  E^{-\Gamma} e^{-(E/E_c)}}$. Since the power-law
index and the e-folding energy E$_c$ are highly correlated we fixed
the photon index to 2.0 (see below). The best fit e-folding energy
is 110.8$^{+68.4}_{-33.0}$\,keV (90\% C.L.) with a reduced chi-squared which is
substantially better than the one of the power-law model (0.9 vs. 1.4 and F-test
probability of 0.08).
The presence of a cutoff at $\sim$100\,keV in the X-ray spectra of Seyfert 1
seems also to be confirmed by the analysis of \cite{deluit03}.

\begin{figure}[h!]
\epsscale{0.7}
	\plotone{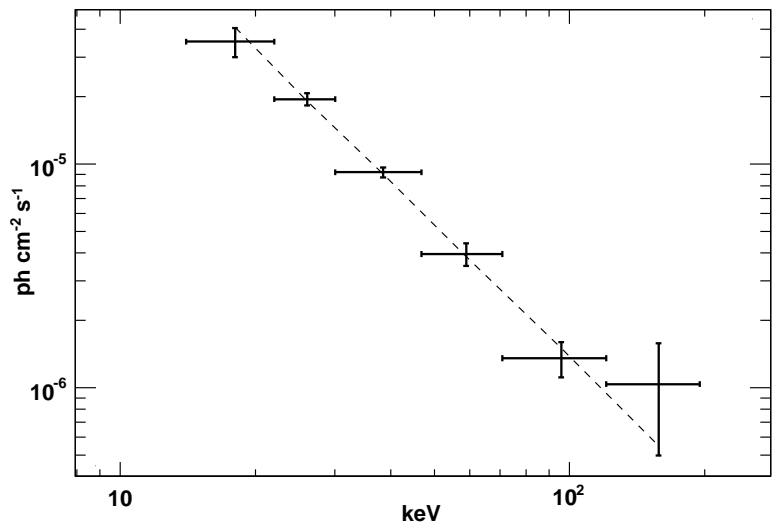} 
	\caption{Stacked spectrum of all AGN reported in Tab. \ref{tab:sample} 
excluding the Blazars. The dashed line is the best power-law fit to the
data (photon index 2.0$\pm0.07$.)}
	\label{fig:stack}
\end{figure}

Finally, we performed the stacked spectral analysis of all the Seyferts to
investigate the averaged spectrum of the local AGN detected by BAT.
The stacked spectrum,
shown in Fig. \ref{fig:stack}, is consistent (in the 15--200\,keV range)
with a power-law model of photon index of 2.00$\pm0.07$ (90\% C.L.).

%
%
\section{The hard X-ray extragalactic sample}\label{sec:extra}
The extragalactic sample, shown in Table \ref{tab:sample},
 was derived from the catalog reported in Table 2
of \cite{ajello07a} considering only objects at $|b|>15^{\circ}$ 
and not spatially associated with the Large Magellanic Cloud.
Here we describe the main properties of the sample.

%
%
\subsection{Completeness of the sample}
\begin{figure}[h!]
\epsscale{0.7}
	\plotone{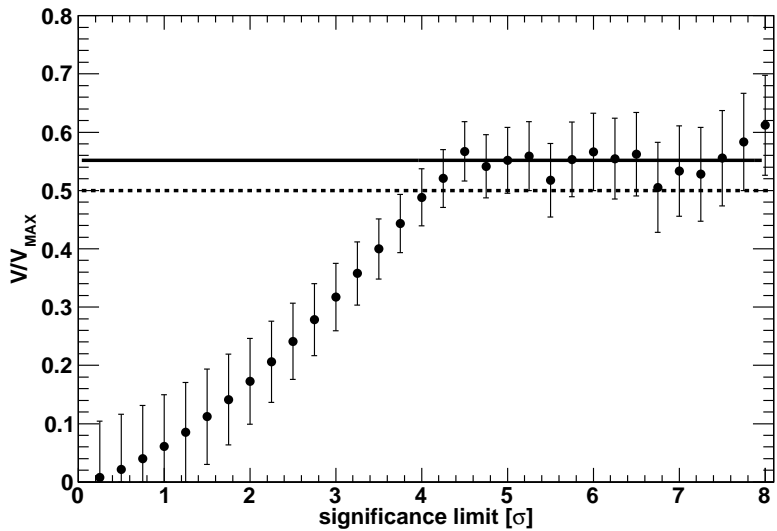} 
	\caption{V/V$_{\rm MAX}$ as a function of detection threshold for
the sample of extragalactic sources. The dashed line is the expected
value  (0.5) for a complete sample in an homogeneous distribution.
The solid line shows the mean test value for S/N$>4.5\sigma$.
}
	\label{fig:vvmax}
\end{figure}

In order to compute the AGN number-flux relation it is necessary to have 
a complete and unbiased sample.
Since different regions of the sky have different exposure times,
we applied
in \cite{ajello07a} a significance limit rather than a flux limit 
to define our sample. Now we want to test our extragalactic sample for 
completeness  (i.e. derive the  significance limit
which ensures to include all objects above a given flux limit) and 
we use the V/V$_{\rm MAX}$ method \citep{schmidt68}\footnotemark{}. 
\footnotetext{
In this test V stands for the volume where the object has been detected
and V$_{\rm MAX}$ is the accessible volume in which the object,
due to the flux limit of the survey, could have been found.
In case of no evolution $<$V/V$_{\rm MAX}$$>$ = 0.5 is expected.}
  This method which is  applied to samples complete to a well-defined 
significance limit, can also be 
 used to test the completeness level of a sample
as a function of significance.
For a significance
limit below the true completeness level limit of the sample, 
the V/V$_{\rm MAX}$ returns a value less than $\langle$V/V$_{\rm MAX}\rangle_{true}$
which would be the true test result for a complete sample. 
Above the completeness
limit the $\langle$V/V$_{\rm MAX}\rangle$ values should be distributed around
$\langle$V/V$_{\rm MAX}\rangle_{true}$ within the statistical uncertainties.
\\\\
V/V$_{\rm MAX}$ is computed for each source as 
$(F/(\sigma_{test}\delta F))^{-3/2}$, where $F$ is the flux,
$\delta F$ is the 1$\sigma$ statistical uncertainty, 
$\sigma_{test}$ is the significance level tested for completeness
(and thus the term $\sigma_{test}\delta F$ is the limiting flux of the sky
region where we detected the source),
and the exponent $-3/2$ comes from the assumption of no evolution
and uniform distribution in the local universe.
$\langle$V/V$_{\rm MAX}\rangle$  is  computed
as an average of all  sources detected with S/N$\geq\sigma_{test}$.
For a given mean value $m=\langle$V/V$_{\rm MAX}\rangle$ and $n$ sources,
the error on $\langle$V/V$_{\rm MAX}\rangle$ can be computed as \citep{avni80}:
\begin{equation}
\sigma_m(n) = \sqrt{ \frac{1/3- m -m^2}{n}} 
\end{equation}
\\\\
The results of the test are shown in Fig. \ref{fig:vvmax}. We find a constant
value for significances $>4.5\sigma$.
The deviation from the expected 0.5 value is insignificant
being less than 1$\sigma$.\footnotemark{} 
\footnotetext{It is not uncommon for coded mask detectors to produce
a test value slightly above 0.5\citep[e.g.][]{beckmann06}. This is likely
caused by systematic errors which tend to increase the
$\langle V/V_{\rm MAX}\rangle$  value.}
\\

We also remark that  for completeness  we are referring to
the  threshold above which all sources above
the corresponding flux limit are included in the sample.
Furthermore, given the small redshift of the sample (see section \ref{sec:stat})
the hypothesis of no evolution is justified.

%
%
%
\subsection{Extragalactic source counts}

\begin{figure}[h!]
\epsscale{0.7}
	\plotone{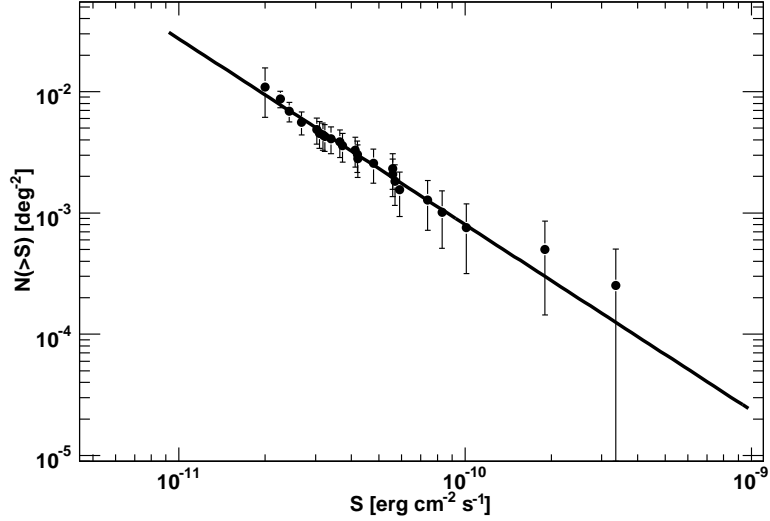} 
	\caption{Extragalactic cumulative source count distribution
 in the 14-170\,keV band.
The solid line is the best fit described in the text.}
	\label{fig:lognlogs}
\end{figure}

The cumulative source number density can be computed as:
\begin{equation}
N(>S)= \sum^{N_{\rm S}}_{i=1} \frac{1}{\Omega_i}\ \ \ \   [{\textup deg^{-2}}]
\end{equation}

where $N_{\rm S}$ is the total number of detected sources in the field with fluxes 
greater than S and $\Omega_i$ is the sky coverage associated to the flux 
of the $i^{th}$ source (shown in  Fig.
9 of \cite{ajello07a}).\\
The cumulative distribution is reported in Fig. \ref{fig:lognlogs}.
We performed a maximum likelihood fit to the cumulative counts
assuming a simple power-law model of the form $N(>S)=A S^{-\alpha}$.
Here, A is the normalization at $2\times 10^{-11}$\,erg cm$^{-2}$ s$^{-1}$
and $\alpha$ is the slope.
As conventional we used the maximum likelihood estimator
\citep[e.g.][]{crawford70} to determine the best fit values.
The normalization is not a parameter of the fit, 
but  is obtained assuming that the number 
of expected sources from the best fit model is equal 
to the total observed number.
The Poissonian error on the total number of sources provides a reliable
estimate of its error.

In the 14--170\,keV band
the best fit parameter is: $\alpha=1.55\pm0.20$ (with normalization
$9.6 \pm1.9 \times 10^{-3}$ deg$^{-2}$). 
The source count distribution is thus consistent with
a pure Euclidean function ($\alpha=3/2$).
From our data we expect
that the number of all-sky AGN is 396$\pm80$
brighter than  $2 \times 10^{-11}$ erg cm$^{-2}$ s$^{-1}$.
This corresponds to an integrated flux of $\sim 5\times10^{-12}$  
erg cm$^{-2}$ s$^{-1}$ deg$^{-2}$ or $\sim$1.5\% of the intensity of the 
X-ray background in the 14-170\,keV energy band as measured by HEAO-1 \citep{gruber99}.\\
We can compare the surface density of extragalactic objects found by BAT
with previous measurements by converting the BAT fluxes to other energy
bands assuming a power law spectrum with photon index of 2.0 
(see Section \ref{sec:spec}) 
and evaluating  the surface density above $10^{-11}$erg cm$^{-2}$ s$^{-1}$.

The results of such comparisons are shown in Tab.\ref{tab:logn_comp}. 
The BAT surface density is in agreement with  the 
reported measurements, except for the case of the 0.5--2 and 2--10\,keV surveys.
Indeed, such surveys, at limiting fluxes of 10$^{-11}$ erg cm$^{-2}$ s$^{-1}$, 
are biased against the detection of absorbed sources\footnotemark{}.
\footnotetext{The bias decreases in deep fields and thus at lower fluxes
(and higher redshifts)
because the photoelectric cut-off is redshifted at lower energies.}
It is also worth noting that the recent XMM measurement of the
5--10\,keV source  counts distribution
\citep{cappelluti07} is in perfect agreement with our estimate.

%
%
\subsection{Statistical properties}\label{sec:stat}
Above the 4.5\,$\sigma$ the extragalactic sample, 
shown in Table \ref{tab:sample}, contains 24 AGN.
19 objects are classified as Seyfert galaxies, 3 as blazars, 1 as 
X-ray bright Optically Normal Galaxy (XBONG) and 1 as Quasar. 
The identification completeness of such sample is thus 100\%.

Excluding the blazars, 
the median redshift of the sample is z=0.026 (mean is z=0.046)
giving a  median luminosity of $10^{43.5}$\,erg s$^{-1}$ 
(mean is $10^{43.8}$\,erg s$^{-1}$)
in the 14-170\,keV band.
Assuming a hydrogen column density of 10$^{22}$ atoms cm$^{-2}$ as the threshold between absorbed and
unabsorbed objects, we find that intrinsic absorption is present 
in $\sim$55\% of the sample.  This fraction is lower than the 75\% 
expected by the standard unified model, which is derived by the opening 
angle of ionizations cones  \citep[e.g.][]{evans91}.
However, this unexpectedly low fraction of absorbed AGN
in the local Universe does not seem to pose any particular problem for the
understanding and the synthesis of the CXB \citep[e.g.][]{sazonov07}.

\begin{figure}[h!]
\epsscale{0.7}
	\plotone{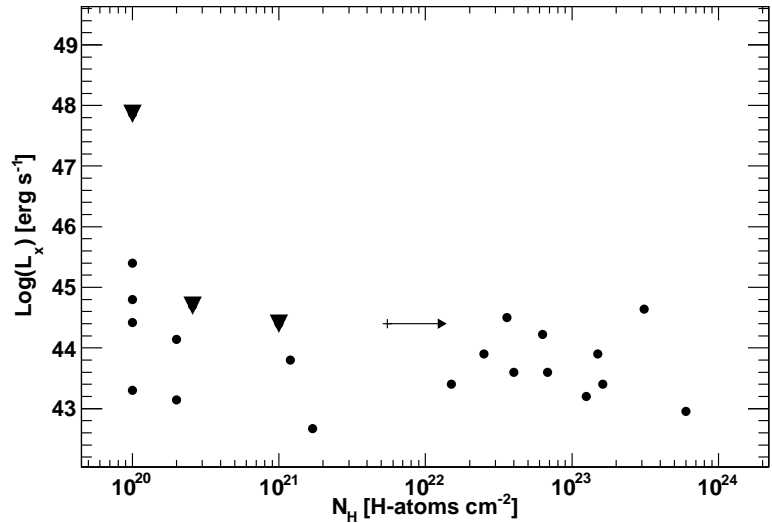} 
	\caption{Luminosity, in the 14-170\,keV band, vs. intrinsic column density
for the extragalactic sample.
The blazars are highlighted with a triangle.}
	\label{fig:nhLx}
\end{figure}

In Fig. \ref{fig:nhLx} we show the intrinsic column density of the sources 
as a function of unabsorbed luminosity in the BAT band. 
Excluding the lower limits
on the absorption, we do not find evidences of an
 anticorrelation between luminosity and  absorption.
We also note the presence of a rare very luminous (L$_x \sim10^{45}$ erg cm$^{-2}$)
highly absorbed (N$_H\sim 10^{23}$atoms cm$^{-2}$) type-2 QSO. 
If the lower limits on the absorption will be confirmed, the
total fraction of such objects might be in the range 5-15\%.

None of the source in Tab.~\ref{tab:sample}, having 2--10\,keV measurement, 
is a Compton-thick AGN. Our claim is supported by several evidences:
\begin{itemize}
\item As shown by \cite{matt97}, for the case of NGC 1068, the spectra of Compton-thick
AGN might be reflection-dominated (i.e. the reflection component is larger
than the trasnmitted one). We thus tried to fit to each source a pure reflection
model (PEXRAV  in Xspec). For all the sources, except Mrk 704, the fit is
statistically unacceptable. However, Mrk 704 is not a Compton-thick source
as \cite{landi07} have recently shown.
\item Compton-thick sources generally show iron lines with
equivalent widths 
of $\sim$1\,keV \citep[e.g.][]{guainazzi05}. The spectral analysis
 (see also values in Tab.~\ref{tab:sample}) 
shows that all sources have iron line equivalent widths smaller than 1\,keV.
\item The thickness parameter T, defined as 
L$_{2-10\,{\textup keV}}$/ L$_{\textup {OIII}}$
\citep[see also][]{bassani99}, can be used to identify Compton-thick sources
(characterized by T$\leq$1). We computed the thickness parameter
for all sources having OIII flux measurements \citep{rau07} 
and 2--10\,keV observations (see Tab.~\ref{tab:sample}).
All sources except NGC~2992 (which is however unabsorbed) 
have thickness parameter values consistent with the values expected
for Compton-thin AGN.
\end{itemize}

We evaluated the radio-loudness of AGN using the R-index  defined
in \cite{laor00} as 
R$\equiv f_{\nu}(5\ \textup{ GHz})/f_{\nu}(4400 \textup{\AA})$; the 
distribution of R-values has been shown to be bimodal with a minimum at R=10,
commonly used to define radio-loud (above 10) versus radio-quiet objects.
Interestingly, we note that
a relevant fraction ($\sim40$\%)  of the BAT AGN is radio-loud and 
that these objects show a systematically harder X-ray spectra 
than Seyfert galaxies (mean of 1.66 vs. 2.00).
There is large consensus that  radio-loud quasars host more massive
black holes than radio-quiet ones \citep[e.g.][]{metcalf06,mclure04}.
However, there is no simple explanation for this radio-loudness dichotomy.
Recently \cite{sikora07} showed that the radio-loudness parameter
inversely correlates with the Eddington ratio 
(fraction of bolometric to Eddington luminosity) for both spiral/disk and 
elliptical galaxies. The fact that spiral-hosted AGN are radio-quiet at
high accretion luminosities supports the idea that the black hole spin plays
a major role in the jet production \citep{sikora07}\footnotemark.
\footnotetext{ In fact, by merging
processes, black holes in elliptical galaxies are expected to have larger
spins than those in spiral/disk galaxies.}
As a confirmation, we find a good correlation of intrinsic 
X-ray luminosity and radio-loudness
(Spearman rank test being 0.57 with probability of 0.003).
Such correlation is expected if there is a fundamental connection between
accretion and jet activity \citep{merloni03}.

%
%
\section{Discussion}\label{sec:disc}

We have used the BAT X-ray survey to study key properties
of the local (z$\leq$0.1) AGN population. Our survey is based on the 
14-170\,keV fluxes and it is  sensitive to AGN with column
densities up to N$_{H}\sim 5\times 10^{24}$ atoms cm$^{-2}$. Indeed, for
a typical source with photon index of 2, the decrease in flux 
for  column densities of N$_{H}\sim 10^{24}$ atoms cm$^{-2}$
is only  $\sim7$\% and $\sim55$\%\footnotemark{}  for 
column densities of 
N$_{H}\sim3\times 10^{24}$ atoms cm$^{-2}$ 
\footnotetext{Photoelectric absorption as well as Compton scattering has 
been taken into account in this estimate.}.
Thus, we can affirm that this survey is relatively unbiased with
respect to photoelectric absorption.


Most of the population synthesis models  \citep{ueda03,treister05,gilli07}
predict that Compton-thick AGN (logN$_H>$24) provide a significant 
contribution to the bulk of the CXB emission at 30\,keV\citep{marshall80}.
Although studies of the local Universe (e.g. Risaliti et al., 1999) 
have shown that Compton-thick objects should be as numerous 
as moderately obscured
AGNs (logN$_H<$24) and thus roughly 1/3 of the total AGN population, 
only a handful of these sources are known \citep{comastri04}. 
\cite{gilli07} estimate that the expected fraction 
of Compton-thick objects at limiting fluxes
probed by BAT and INTEGRAL ($\sim10^{-11}$erg cm$^{-2}$ s$^{-1}$)
is in the 15--20\% range. However  the  measured 
fraction of detected Compton-thick objects by these instruments
is, so far, close to, or less than,  10\% \citep{markwardt05,beckmann06}.

The BAT extragalactic sample contains only one source
SWIFT J0823.4-0457, which given its colors (see Fig.~\ref{fig:HR})
might be Compton-thick. However, the joint XRT and BAT spectra
show that the absorption is below the Compton-thick level 
(N$_H\sim 10^{23}$\,atoms cm$^{-2}$). We must therefore conclude that no
Compton-thick AGN are present in our extragalactic sample. 
The probability of not detecting  Compton-thick objects in a sample
of 24 AGN when the expected fraction is 20\% (15\%) is $\sim$0.007 ($\sim$0.03)
while it is 0.1 if the expected fraction is 10\%. 
These probabilities increase 
(0.03, 0.09 and 0.2 for the 20\%, 15\% and 10\% cases)
 if we assume that the only source which lacks
$<10$\,keV measurement (J0854.7+1502) is Compton-thick. Thus, the BAT data
discard tat $>$2\,$\sigma$ level
the hypothesis that Compton-thick AGN may represent a fraction
of $\sim20$\% of the total AGN population.

We find that Sy2s have harder spectra than Sy1s in agreement with what
has been deduced from  {\it OSSE}, BeppoSAX   and INTEGRAL data
\citep[][respectively]{zdziarski00,deluit03,beckmann06}. We tested whether this difference could
be accounted for  by Compton reflection and/or by a high-energy cut-off.
We find that the reflection component improves the fit to the Sy1 averaged
spectrum
(the F-test shows that the reflection is significant at more than the 92\%
level), but it leaves unaltered the photon index. Thus, the difference
in photon indices among Sy1s and Sy2s cannot be ascribed solely
to orientation effects (a stronger reflection is expected for face-on
objects). The spectra of Sy1s show hints of a spectral cut-off 
at $\sim$100\,keV in agreement with \cite{deluit03}. 
According to thermal Compton models, the absence of a cut-off in Sy2s might
indicate a higher temperature of the Comptonizing medium 
(with respect to Sy1s) or that non-thermal Compton scattering plays an
important role.
Nevertheless, given the low S/N of our sources evidences for the cut-off
in the Sy1 spectra are weak.

The best power-law fit to the extragalactic source counts distribution 
yields a slope of $\alpha=1.55\pm0.20$ which is 
 consistent with an Euclidean distribution.
From the best fit,
we derive a surface density of AGN of 
$9.6 \pm1.9 \times 10^{-3}$\,deg$^{-2}$
above the flux limit of 2$\times 10^{-11}$\,erg cm$^{-2}$ s$^{-1}$; 
this estimate is in
very good agreement, when converted to the 20-40\,keV band, with the recently 
derived  source counts distribution  based on INTEGRAL data \citep{beckmann06}.
\cite{beckmann06} find a slope of 1.66$\pm0.11$ which is also
consistent with our measurement, but steeper than the 1.5 Euclidean
value. Even though this could be due to a non-perfectly computed sky coverage,
the authors suggest that the distribution of AGNs in the local Universe may not
be isotropic because of local clustering of sources 
(e.g. the local group of galaxies).

The BAT source count distribution resolves only 1-2\% of the CXB
into extragalactic sources; nevertheless as it is unbiased
with respect to absorption
it gives important information relative to the fraction of obscured
sources which are missed by deep $<10$keV surveys because of absorption.
The extrapolation of the BAT source count distribution to the 
2-10\,keV band assuming an unabsorbed spectrum with photon index 2
yields a surface density of AGN of 1.6$\pm0.32 \times10^{-2}$deg$^{-2}$ above
10$^{-11}$erg cm$^{-2}$ s$^{-1}$; while the surface density as extrapolated 
to brighter fluxes by XMM \citep{cappelluti07} and as 
predicted by the model of \cite{gilli07} is 0.9$\times10^{-2}$deg$^{-2}$.
The factor $\sim2$ more sources BAT sees 
 can be explained in term of absorption. Indeed, if we
take into account the absorption distribution derived
for BAT AGNs by \cite{markwardt05}
 (thus assuming that 66\% of all AGN are
absorbed with a mean column density of $10^{23}$ atoms cm$^{-2}$)
we get a surface density 0.86$\pm0.17\times 10^{-2}$deg$^{-2}$
which is consistent with the XMM extrapolation and the model prediction.

The extragalactic sample is composed of $\sim$90\% emission-line galaxies
and $\sim$10\% blazars. We find that 55\% of the emission-line galaxies
are obscured by absorbing columns larger than 10$^{22}$\,H-atoms cm$^{-2}$.
This fraction is in agreement with the INTEGRAL measurements 
\citep[e.g.][]{sazonov07}, but less
than what is suggested ($\sim75$\%) by the unified AGN model.
However, \cite{sazonov07} successfully showed that low-luminosity
(mostly absorbed) AGN account for much, $\sim$90\%, 
of the luminosity density of the  local Universe.
This is also confirmed by the model of \cite{gilli07} which shows that
the required fraction of obscured sources varies with intrinsic luminosity
being 3.7 and 1.0 below and above $10^{43.5}$\,erg s$^{-1}$.
A relevant fraction ($\sim$40\%) of the BAT-detected AGN is radio-loud.
These objects show a systematically harder X-ray spectra 
than Seyfert galaxies (1.66 vs. 2.00). The hard photon index and the
correlation of radio-loudness with X-ray luminosity suggest that
a jet is presently at work in all these objects.
Our sample also comprises 1 (and possibly up to 3 considering the ROSAT
lower limits on the absorption) highly luminous highly absorbed QSO.

%
%

\section{Summary}
We use the {\it Swift}/BAT instrument to study the properties of
the local (z$\leq$1) AGN in connection with the synthesis of the X-ray
background emission.
The results of this study can be summarized as follows:
\begin{itemize}
\item Despite the consensus that Compton-thick objects may represent
a substantial fraction of the local AGN population
\citep[e.g.][]{risaliti99,gilli07},
we do not detect any such object. The probability associated to this
non-detection is 0.007, 0.03 and 0.1 when assuming that their fraction
should be 20\%, 15\% and 10\% of the total AGNs.
BAT discards at $>2$\,$\sigma$  the hypothesis that the fraction
of Compton-thick objects is 20\%.

\item Seyfert 2 galaxies have harder X-ray 
spectra than Seyfert 1. We find that
this difference cannot be ascribed solely to the different viewing angle
and thus to the different amount of Compton reflection which is expected.
The Seyfert 1 galaxies comprised in our sample 
show weak evidences  for a spectral cut-off in the 
$\sim$100\,keV range. This might highlight an intrinsic difference
among the two classes. Indeed, the absence of a cut-off 
in the spectra of Seyfert 2s
might indicate a different (higher) temperature of the Comptonizing medium
or that non-thermal Compton scattering play an important role.

\item The best power-law fit to the extragalactic source counts is consistent
with a Euclidean function with slope of 1.55$\pm0.20$. At the current limiting
fluxes (2$\times10^{-11}$\,erg cm$^{-2}$ s$^{-1}$), BAT resolves only 1--2\%
of the CXB emission in the 14--170\,keV band.

\item The fraction of emission-line AGN which is absorbed by 
N$_H>10^{22}$\,atoms cm$^{-2}$ is $\sim$55\%. This is lower than
the 75\% expected by the standard AGN unified model.

\end{itemize}

This work shows the capabilities of BAT to produce an unbiased sample 
of AGN which is important for the understanding of the synthesis
of the CXB emission in the hard X-ray band.


\acknowledgments
MA acknowledges N. Gehrels and the BAT team for hospitality,
M. Capalbi for assistance during Beppo-SAX and Swift-XRT  data analysis, 
N. Cappelluti for useful discussions on the source count 
distribution derivation, R. Mushotzky for valuable suggestions and the 
anonymous referee for his comments which helped improving the paper.
This research has made use of the NASA/IPAC extragalactic 
Database (NED) which
is operated by the Jet Propulsion Laboratory, of data obtained from the 
High Energy Astrophysics Science Archive Research Center (HEASARC) provided 
by NASA's Goddard Space Flight Center, of the SIMBAD Astronomical Database
which is operated by the Centre de Donn\'ees astronomiques de Strasbourg, of
the Sloan Digital Sky Survey (SDSS) managed by the Astrophysical Research 
Consortium (ARC) for the Participating Institutions and of the ROSAT All Sky
Survey mantained by the Max Planck Institut f\"ur Extraterrestrische Physik. 

%
%
\bibliographystyle{apj}
\bibliography{/Users/marcoajello/Work/Papers/BiblioLib/biblio}

\begin{thebibliography}{59}
\expandafter\ifx\csname natexlab\endcsname\relax\def\natexlab#1{#1}\fi

\bibitem[{{Ajello} {et~al.}(2007){Ajello}, {Greiner}, {Kanbach}, {Rau},
  {Strong}, \& {Kennea}}]{ajello07a}
{Ajello}, M., {Greiner}, J., {Kanbach}, G., {Rau}, A., {Strong}, A.~W., \&
  {Kennea}, J.~A. 2007, ApJ, submitted

\bibitem[{{Ajello} {et~al.}(2006){Ajello}, {Greiner}, {Rau}, {Barthelmy},
  {Kennea}, {Falcone}, {Godet}, {Grupe}, {Tueller}, {Markwardt}, {Mushotsky},
  {Belloni}, {Mukai}, {Holland}, \& {Gehrels}}]{ajello06a}
{Ajello}, M., {Greiner}, J., {Rau}, A., {Barthelmy}, S., {Kennea}, J.~A.,
  {Falcone}, A., {Godet}, O., {Grupe}, D., {Tueller}, J., {Markwardt}, C.,
  {Mushotsky}, R., {Belloni}, T., {Mukai}, K., {Holland}, S.~T., \& {Gehrels},
  N. 2006, The Astronomer's Telegram, 697, 1

\bibitem[{{Antonucci}(1993)}]{antonucci93}
{Antonucci}, R. 1993, \araa, 31, 473

\bibitem[{{Arnaud}(1996)}]{arnaud96}
{Arnaud}, K.~A. 1996, in Astronomical Society of the Pacific Conference Series,
  Vol. 101, Astronomical Data Analysis Software and Systems V, ed. G.~H.
  {Jacoby} \& J.~{Barnes}, 17--+

\bibitem[{{Avni} \& {Bahcall}(1980)}]{avni80}
{Avni}, Y. \& {Bahcall}, J.~N. 1980, \apj, 235, 694

\bibitem[{{Barthelmy} {et~al.}(2005){Barthelmy}, {Barbier}, {Cummings},
  {Fenimore}, {Gehrels}, {Hullinger}, {Krimm}, {Markwardt}, {Palmer},
  {Parsons}, {Sato}, {Suzuki}, {Takahashi}, {Tashiro}, \&
  {Tueller}}]{barthelmy05}
{Barthelmy}, S.~D., {Barbier}, L.~M., {Cummings}, J.~R., {Fenimore}, E.~E.,
  {Gehrels}, N., {Hullinger}, D., {Krimm}, H.~A., {Markwardt}, C.~B., {Palmer},
  D.~M., {Parsons}, A., {Sato}, G., {Suzuki}, M., {Takahashi}, T., {Tashiro},
  M., \& {Tueller}, J. 2005, Space Science Reviews, 120, 143

\bibitem[{{Bassani} {et~al.}(1999){Bassani}, {Dadina}, {Maiolino}, {Salvati},
  {Risaliti}, {della Ceca}, {Matt}, \& {Zamorani}}]{bassani99}
{Bassani}, L., {Dadina}, M., {Maiolino}, R., {Salvati}, M., {Risaliti}, G.,
  {della Ceca}, R., {Matt}, G., \& {Zamorani}, G. 1999, \apjs, 121, 473

\bibitem[{{Beckmann} {et~al.}(2006){Beckmann}, {Soldi}, {Shrader}, {Gehrels},
  \& {Produit}}]{beckmann06}
{Beckmann}, V., {Soldi}, S., {Shrader}, C.~R., {Gehrels}, N., \& {Produit}, N.
  2006, \apj, 652, 126

\bibitem[{{Bird} {et~al.}(2006){Bird}, {Barlow}, {Bassani}, {Bazzano},
  {B{\'e}langer}, {Bodaghee}, {Capitanio}, {Dean}, {Fiocchi}, {Hill}, {Lebrun},
  {Malizia}, {Mas-Hesse}, {Molina}, {Moran}, {Renaud}, {Sguera}, {Shaw},
  {Stephen}, {Terrier}, {Ubertini}, {Walter}, {Willis}, \& {Winkler}}]{bird06}
{Bird}, A.~J., {Barlow}, E.~J., {Bassani}, L., {Bazzano}, A., {B{\'e}langer},
  G., {Bodaghee}, A., {Capitanio}, F., {Dean}, A.~J., {Fiocchi}, M., {Hill},
  A.~B., {Lebrun}, F., {Malizia}, A., {Mas-Hesse}, J.~M., {Molina}, M.,
  {Moran}, L., {Renaud}, M., {Sguera}, V., {Shaw}, S.~E., {Stephen}, J.~B.,
  {Terrier}, R., {Ubertini}, P., {Walter}, R., {Willis}, D.~R., \& {Winkler},
  C. 2006, \apj, 636, 765

\bibitem[{{Bolton} \& {Butler}(1975)}]{bolton75}
{Bolton}, J.~G. \& {Butler}, P.~W. 1975, Australian Journal of Physics
  Astrophysical Supplement, 34, 33

\bibitem[{{Cappelluti} {et~al.}(2007){Cappelluti}, {Hasinger}, {Brusa},
  {Comastri}, {Zamorani}, {Boehringer}, {Brunner}, {Civano}, {Finoguenov},
  {Fiore}, {Gilli}, {Griffiths}, {Mainieri}, {Matute}, {Miyaji}, \&
  {Silverman}}]{cappelluti07}
{Cappelluti}, N., {Hasinger}, G., {Brusa}, M., {Comastri}, A., {Zamorani}, G.,
  {Boehringer}, H., {Brunner}, H., {Civano}, F., {Finoguenov}, A., {Fiore}, F.,
  {Gilli}, R., {Griffiths}, R.~E., {Mainieri}, V., {Matute}, I., {Miyaji}, T.,
  \& {Silverman}, J. 2007, ArXiv e-prints, 704

\bibitem[{{Comastri}(2004)}]{comastri04}
{Comastri}, A. 2004, in Astrophysics and Space Science Library, Vol. 308,
  Astrophysics and Space Science Library, ed. A.~J. {Barger}, 245--+

\bibitem[{{Comastri} {et~al.}(1995){Comastri}, {Setti}, {Zamorani}, \&
  {Hasinger}}]{comastri95}
{Comastri}, A., {Setti}, G., {Zamorani}, G., \& {Hasinger}, G. 1995, \aap, 296,
  1

\bibitem[{{Crawford} \& {Fabian}(1995)}]{crawford95}
{Crawford}, C.~S. \& {Fabian}, A.~C. 1995, \mnras, 273, 827

\bibitem[{{Crawford} {et~al.}(1970){Crawford}, {Jauncey}, \&
  {Murdoch}}]{crawford70}
{Crawford}, D.~F., {Jauncey}, D.~L., \& {Murdoch}, H.~S. 1970, \apj, 162, 405

\bibitem[{{Deluit} \& {Courvoisier}(2003)}]{deluit03}
{Deluit}, S. \& {Courvoisier}, T.~J.-L. 2003, \aap, 399, 77

\bibitem[{{den Hartog} {et~al.}(2004){den Hartog}, {Hermsen}, {Kuiper}, {in't
  Zand}, {Winkler}, \& {Domingo}}]{denhartog04}
{den Hartog}, P.~R., {Hermsen}, W., {Kuiper}, L.~M., {in't Zand}, J.~J.~M.,
  {Winkler}, C., \& {Domingo}, A. 2004, The Astronomer's Telegram, 261, 1

\bibitem[{{Elvis} {et~al.}(1992){Elvis}, {Plummer}, {Schachter}, \&
  {Fabbiano}}]{elvis92}
{Elvis}, M., {Plummer}, D., {Schachter}, J., \& {Fabbiano}, G. 1992, \apjs, 80,
  257

\bibitem[{{Evans} {et~al.}(1991){Evans}, {Ford}, {Kinney}, {Antonucci},
  {Armus}, \& {Caganoff}}]{evans91}
{Evans}, I.~N., {Ford}, H.~C., {Kinney}, A.~L., {Antonucci}, R.~R.~J., {Armus},
  L., \& {Caganoff}, S. 1991, \apjl, 369, L27

\bibitem[{{Fenimore} \& {Cannon}(1978)}]{fenimore78}
{Fenimore}, E.~E. \& {Cannon}, T.~M. 1978, \ao, 17, 337

\bibitem[{{Gehrels} {et~al.}(2004){Gehrels}, {Chincarini}, {Giommi}, {Mason},
  {Nousek}, {Wells}, {White}, {Barthelmy}, {Burrows}, {Cominsky}, {Hurley},
  {Marshall}, {M{\'e}sz{\'a}ros}, {Roming}, {Angelini}, {Barbier}, {Belloni},
  {Campana}, {Caraveo}, {Chester}, {Citterio}, {Cline}, {Cropper}, {Cummings},
  {Dean}, {Feigelson}, {Fenimore}, {Frail}, {Fruchter}, {Garmire}, {Gendreau},
  {Ghisellini}, {Greiner}, {Hill}, {Hunsberger}, {Krimm}, {Kulkarni}, {Kumar},
  {Lebrun}, {Lloyd-Ronning}, {Markwardt}, {Mattson}, {Mushotzky}, {Norris},
  {Osborne}, {Paczynski}, {Palmer}, {Park}, {Parsons}, {Paul}, {Rees},
  {Reynolds}, {Rhoads}, {Sasseen}, {Schaefer}, {Short}, {Smale}, {Smith},
  {Stella}, {Tagliaferri}, {Takahashi}, {Tashiro}, {Townsley}, {Tueller},
  {Turner}, {Vietri}, {Voges}, {Ward}, {Willingale}, {Zerbi}, \&
  {Zhang}}]{gehrels04}
{Gehrels}, N., {Chincarini}, G., {Giommi}, P., {Mason}, K.~O., {Nousek}, J.~A.,
  {Wells}, A.~A., {White}, N.~E., {Barthelmy}, S.~D., {Burrows}, D.~N.,
  {Cominsky}, L.~R., {Hurley}, K.~C., {Marshall}, F.~E., {M{\'e}sz{\'a}ros},
  P., {Roming}, P.~W.~A., {Angelini}, L., {Barbier}, L.~M., {Belloni}, T.,
  {Campana}, S., {Caraveo}, P.~A., {Chester}, M.~M., {Citterio}, O., {Cline},
  T.~L., {Cropper}, M.~S., {Cummings}, J.~R., {Dean}, A.~J., {Feigelson},
  E.~D., {Fenimore}, E.~E., {Frail}, D.~A., {Fruchter}, A.~S., {Garmire},
  G.~P., {Gendreau}, K., {Ghisellini}, G., {Greiner}, J., {Hill}, J.~E.,
  {Hunsberger}, S.~D., {Krimm}, H.~A., {Kulkarni}, S.~R., {Kumar}, P.,
  {Lebrun}, F., {Lloyd-Ronning}, N.~M., {Markwardt}, C.~B., {Mattson}, B.~J.,
  {Mushotzky}, R.~F., {Norris}, J.~P., {Osborne}, J., {Paczynski}, B.,
  {Palmer}, D.~M., {Park}, H.-S., {Parsons}, A.~M., {Paul}, J., {Rees}, M.~J.,
  {Reynolds}, C.~S., {Rhoads}, J.~E., {Sasseen}, T.~P., {Schaefer}, B.~E.,
  {Short}, A.~T., {Smale}, A.~P., {Smith}, I.~A., {Stella}, L., {Tagliaferri},
  G., {Takahashi}, T., {Tashiro}, M., {Townsley}, L.~K., {Tueller}, J.,
  {Turner}, M.~J.~L., {Vietri}, M., {Voges}, W., {Ward}, M.~J., {Willingale},
  R., {Zerbi}, F.~M., \& {Zhang}, W.~W. 2004, \apj, 611, 1005

\bibitem[{{Giacconi} {et~al.}(1962){Giacconi}, {Gursky}, {Paolini}, \&
  {Rossi}}]{giacconi62}
{Giacconi}, R., {Gursky}, H., {Paolini}, F.~R., \& {Rossi}, B.~B. 1962,
  Physical Review Letters, 9, 439

\bibitem[{{Gilli} {et~al.}(2007){Gilli}, {Comastri}, \& {Hasinger}}]{gilli07}
{Gilli}, R., {Comastri}, A., \& {Hasinger}, G. 2007, \aap, 463, 79

\bibitem[{{Gilli} {et~al.}(2001){Gilli}, {Salvati}, \& {Hasinger}}]{gilli01}
{Gilli}, R., {Salvati}, M., \& {Hasinger}, G. 2001, \aap, 366, 407

\bibitem[{{Giommi} {et~al.}(1991){Giommi}, {Tagliaferri}, {Beuermann},
  {Branduardi-Raymont}, {Brissenden}, {Graser}, {Mason}, {Mittaz}, {Murdin},
  {Pooley}, {Thomas}, \& {Tuohy}}]{giommi91}
{Giommi}, P., {Tagliaferri}, G., {Beuermann}, K., {Branduardi-Raymont}, G.,
  {Brissenden}, R., {Graser}, U., {Mason}, K.~O., {Mittaz}, J.~D.~P., {Murdin},
  P., {Pooley}, G., {Thomas}, H.-C., \& {Tuohy}, I. 1991, \apj, 378, 77

\bibitem[{{G{\"o}tz} {et~al.}(2006){G{\"o}tz}, {Mereghetti}, {Merlini},
  {Sidoli}, \& {Belloni}}]{gotz06}
{G{\"o}tz}, D., {Mereghetti}, S., {Merlini}, D., {Sidoli}, L., \& {Belloni}, T.
  2006, \aap, 448, 873

\bibitem[{{Gruber} {et~al.}(1999){Gruber}, {Matteson}, {Peterson}, \&
  {Jung}}]{gruber99}
{Gruber}, D.~E., {Matteson}, J.~L., {Peterson}, L.~E., \& {Jung}, G.~V. 1999,
  \apj, 520, 124

\bibitem[{{Guainazzi} {et~al.}(2005){Guainazzi}, {Matt}, \&
  {Perola}}]{guainazzi05}
{Guainazzi}, M., {Matt}, G., \& {Perola}, G.~C. 2005, \aap, 444, 119

\bibitem[{{Jonker} {et~al.}(2001){Jonker}, {van der Klis}, {Homan},
  {M{\'e}ndez}, {van Paradijs}, {Belloni}, {Kouveliotou}, {Lewin}, \&
  {Ford}}]{jonker01}
{Jonker}, P.~G., {van der Klis}, M., {Homan}, J., {M{\'e}ndez}, M., {van
  Paradijs}, J., {Belloni}, T., {Kouveliotou}, C., {Lewin}, W., \& {Ford},
  E.~C. 2001, \apj, 553, 335

\bibitem[{{Landi} {et~al.}(2007){Landi}, {Masetti}, {Morelli}, {Palazzi},
  {Bassani}, {Malizia}, {Bazzano}, {Bird}, {Dean}, {Galaz}, {Minniti}, \&
  {Ubertini}}]{landi07}
{Landi}, R., {Masetti}, N., {Morelli}, L., {Palazzi}, E., {Bassani}, L.,
  {Malizia}, A., {Bazzano}, A., {Bird}, A.~J., {Dean}, A.~J., {Galaz}, G.,
  {Minniti}, D., \& {Ubertini}, P. 2007, ArXiv e-prints, 706

\bibitem[{{Laor}(2000)}]{laor00}
{Laor}, A. 2000, \apjl, 543, L111

\bibitem[{{Magdziarz} \& {Zdziarski}(1995)}]{magdziarz95}
{Magdziarz}, P. \& {Zdziarski}, A.~A. 1995, \mnras, 273, 837

\bibitem[{{Malizia} {et~al.}(2002){Malizia}, {Malaguti}, {Bassani}, {Cappi},
  {Comastri}, {Di Cocco}, {Palazzi}, \& {Vignali}}]{malizia02}
{Malizia}, A., {Malaguti}, G., {Bassani}, L., {Cappi}, M., {Comastri}, A., {Di
  Cocco}, G., {Palazzi}, E., \& {Vignali}, C. 2002, \aap, 394, 801

\bibitem[{{Markwardt} {et~al.}(2005){Markwardt}, {Tueller}, {Skinner},
  {Gehrels}, {Barthelmy}, \& {Mushotzky}}]{markwardt05}
{Markwardt}, C.~B., {Tueller}, J., {Skinner}, G.~K., {Gehrels}, N.,
  {Barthelmy}, S.~D., \& {Mushotzky}, R.~F. 2005, \apjl, 633, L77

\bibitem[{{Marshall} {et~al.}(1980){Marshall}, {Boldt}, {Holt}, {Miller},
  {Mushotzky}, {Rose}, {Rothschild}, \& {Serlemitsos}}]{marshall80}
{Marshall}, F.~E., {Boldt}, E.~A., {Holt}, S.~S., {Miller}, R.~B., {Mushotzky},
  R.~F., {Rose}, L.~A., {Rothschild}, R.~E., \& {Serlemitsos}, P.~J. 1980,
  \apj, 235, 4

\bibitem[{{Masetti} {et~al.}(2006{\natexlab{a}}){Masetti}, {Bassani}, {Dean},
  {Ubertini}, \& {Walter}}]{masetti06}
{Masetti}, N., {Bassani}, L., {Dean}, A.~J., {Ubertini}, P., \& {Walter}, R.
  2006{\natexlab{a}}, The Astronomer's Telegram, 735, 1

\bibitem[{{Masetti} {et~al.}(2006{\natexlab{b}}){Masetti}, {Morelli},
  {Palazzi}, {Galaz}, {Bassani}, {Bazzano}, {Bird}, {Dean}, {Israel}, {Landi},
  {Malizia}, {Minniti}, {Schiavone}, {Stephen}, {Ubertini}, \&
  {Walter}}]{masetti06b}
{Masetti}, N., {Morelli}, L., {Palazzi}, E., {Galaz}, G., {Bassani}, L.,
  {Bazzano}, A., {Bird}, A.~J., {Dean}, A.~J., {Israel}, G.~L., {Landi}, R.,
  {Malizia}, A., {Minniti}, D., {Schiavone}, F., {Stephen}, J.~B., {Ubertini},
  P., \& {Walter}, R. 2006{\natexlab{b}}, \aap, 459, 21

\bibitem[{{Matt} {et~al.}(1997){Matt}, {Guainazzi}, {Frontera}, {Bassani},
  {Brandt}, {Fabian}, {Fiore}, {Haardt}, {Iwasawa}, {Maiolino}, {Malaguti},
  {Marconi}, {Matteuzzi}, {Molendi}, {Perola}, {Piraino}, \& {Piro}}]{matt97}
{Matt}, G., {Guainazzi}, M., {Frontera}, F., {Bassani}, L., {Brandt}, W.~N.,
  {Fabian}, A.~C., {Fiore}, F., {Haardt}, F., {Iwasawa}, K., {Maiolino}, R.,
  {Malaguti}, G., {Marconi}, A., {Matteuzzi}, A., {Molendi}, S., {Perola},
  G.~C., {Piraino}, S., \& {Piro}, L. 1997, \aap, 325, L13

\bibitem[{{Matt} {et~al.}(2003){Matt}, {Guainazzi}, \& {Maiolino}}]{matt03}
{Matt}, G., {Guainazzi}, M., \& {Maiolino}, R. 2003, \mnras, 342, 422

\bibitem[{{McLure} \& {Jarvis}(2004)}]{mclure04}
{McLure}, R.~J. \& {Jarvis}, M.~J. 2004, \mnras, 353, L45

\bibitem[{{Merloni} {et~al.}(2003){Merloni}, {Heinz}, \& {di
  Matteo}}]{merloni03}
{Merloni}, A., {Heinz}, S., \& {di Matteo}, T. 2003, \mnras, 345, 1057

\bibitem[{{Metcalf} \& {Magliocchetti}(2006)}]{metcalf06}
{Metcalf}, R.~B. \& {Magliocchetti}, M. 2006, \mnras, 365, 101

\bibitem[{{Norton} {et~al.}(2000){Norton}, {Beardmore}, {Retter}, \&
  {Buckley}}]{norton00}
{Norton}, A.~J., {Beardmore}, A.~P., {Retter}, A., \& {Buckley}, D.~A.~H. 2000,
  \mnras, 312, 362

\bibitem[{{Rau} {et~al.}(2007){Rau}, {Greiner}, {Salvato}, \& {Ajello}}]{rau07}
{Rau}, A., {Greiner}, J., {Salvato}, M., \& {Ajello}, M. 2007, A\&A, submitted

\bibitem[{{Revnivtsev} {et~al.}(2006){Revnivtsev}, {Sazonov}, {Churazov}, \&
  {Trudolyubov}}]{revnivtsev06}
{Revnivtsev}, M., {Sazonov}, S., {Churazov}, E., \& {Trudolyubov}, S. 2006,
  \aap, 448, L49

\bibitem[{{Risaliti} {et~al.}(1999){Risaliti}, {Maiolino}, \&
  {Salvati}}]{risaliti99}
{Risaliti}, G., {Maiolino}, R., \& {Salvati}, M. 1999, \apj, 522, 157

\bibitem[{{Sazonov} {et~al.}(2007){Sazonov}, {Revnivtsev}, {Krivonos},
  {Churazov}, \& {Sunyaev}}]{sazonov07}
{Sazonov}, S., {Revnivtsev}, M., {Krivonos}, R., {Churazov}, E., \& {Sunyaev},
  R. 2007, \aap, 462, 57

\bibitem[{{Schmidt}(1968)}]{schmidt68}
{Schmidt}, M. 1968, \apj, 151, 393

\bibitem[{{Schwope} {et~al.}(2000){Schwope}, {Hasinger}, {Lehmann}, {Schwarz},
  {Brunner}, {Neizvestny}, {Ugryumov}, {Balega}, {Tr{\"u}mper}, \&
  {Voges}}]{schwope00}
{Schwope}, A., {Hasinger}, G., {Lehmann}, I., {Schwarz}, R., {Brunner}, H.,
  {Neizvestny}, S., {Ugryumov}, A., {Balega}, Y., {Tr{\"u}mper}, J., \&
  {Voges}, W. 2000, Astronomische Nachrichten, 321, 1

\bibitem[{{Sikora} {et~al.}(2007){Sikora}, {Stawarz}, \& {Lasota}}]{sikora07}
{Sikora}, M., {Stawarz}, {\L}., \& {Lasota}, J.-P. 2007, \apj, 658, 815

\bibitem[{{Slowikowska} {et~al.}(2006){Slowikowska}, {Kanbach}, {Borkowski}, \&
  {Becker}}]{slowikowska06}
{Slowikowska}, A., {Kanbach}, G., {Borkowski}, J., \& {Becker}, W. 2006, On the
  Present and Future of Pulsar Astronomy, 26th meeting of the IAU, Joint
  Discussion 2, 16-17 August, 2006, Prague, Czech Republic, JD02, \#8, 2

\bibitem[{{Sowards-Emmerd} {et~al.}(2004){Sowards-Emmerd}, {Romani},
  {Michelson}, \& {Ulvestad}}]{sowards04}
{Sowards-Emmerd}, D., {Romani}, R.~W., {Michelson}, P.~F., \& {Ulvestad}, J.~S.
  2004, \apj, 609, 564

\bibitem[{{Treister} \& {Urry}(2005)}]{treister05}
{Treister}, E. \& {Urry}, C.~M. 2005, \apj, 630, 115

\bibitem[{{Tueller} {et~al.}(2005){Tueller}, {Barthelmy}, {Burrows}, {Falcone},
  {Gehrels}, {Grupe}, {Kennea}, {Markwardt}, {Mushotzky}, \&
  {Skinner}}]{tueller05b}
{Tueller}, J., {Barthelmy}, S., {Burrows}, D., {Falcone}, A., {Gehrels}, N.,
  {Grupe}, D., {Kennea}, J., {Markwardt}, C.~B., {Mushotzky}, R.~F., \&
  {Skinner}, G.~K. 2005, The Astronomer's Telegram, 669, 1

\bibitem[{{Ueda} {et~al.}(2003){Ueda}, {Akiyama}, {Ohta}, \& {Miyaji}}]{ueda03}
{Ueda}, Y., {Akiyama}, M., {Ohta}, K., \& {Miyaji}, T. 2003, \apj, 598, 886

\bibitem[{{Voges} {et~al.}(1999){Voges}, {Aschenbach}, {Boller},
  {Br{\"a}uninger}, {Briel}, {Burkert}, {Dennerl}, {Englhauser}, {Gruber},
  {Haberl}, {Hartner}, {Hasinger}, {K{\"u}rster}, {Pfeffermann}, {Pietsch},
  {Predehl}, {Rosso}, {Schmitt}, {Tr{\"u}mper}, \& {Zimmermann}}]{voges99}
{Voges}, W., {Aschenbach}, B., {Boller}, T., {Br{\"a}uninger}, H., {Briel}, U.,
  {Burkert}, W., {Dennerl}, K., {Englhauser}, J., {Gruber}, R., {Haberl}, F.,
  {Hartner}, G., {Hasinger}, G., {K{\"u}rster}, M., {Pfeffermann}, E.,
  {Pietsch}, W., {Predehl}, P., {Rosso}, C., {Schmitt}, J.~H.~M.~M.,
  {Tr{\"u}mper}, J., \& {Zimmermann}, H.~U. 1999, \aap, 349, 389

\bibitem[{{Worsley} {et~al.}(2005){Worsley}, {Fabian}, {Bauer}, {Alexander},
  {Hasinger}, {Mateos}, {Brunner}, {Brandt}, \& {Schneider}}]{worsley05}
{Worsley}, M.~A., {Fabian}, A.~C., {Bauer}, F.~E., {Alexander}, D.~M.,
  {Hasinger}, G., {Mateos}, S., {Brunner}, H., {Brandt}, W.~N., \& {Schneider},
  D.~P. 2005, \mnras, 357, 1281

\bibitem[{{Zamorani} {et~al.}(1981){Zamorani}, {Henry}, {Maccacaro},
  {Tananbaum}, {Soltan}, {Avni}, {Liebert}, {Stocke}, {Strittmatter},
  {Weymann}, {Smith}, \& {Condon}}]{zamorani81}
{Zamorani}, G., {Henry}, J.~P., {Maccacaro}, T., {Tananbaum}, H., {Soltan}, A.,
  {Avni}, Y., {Liebert}, J., {Stocke}, J., {Strittmatter}, P.~A., {Weymann},
  R.~J., {Smith}, M.~G., \& {Condon}, J.~J. 1981, \apj, 245, 357

\bibitem[{{Zdziarski} {et~al.}(2000){Zdziarski}, {Poutanen}, \&
  {Johnson}}]{zdziarski00}
{Zdziarski}, A.~A., {Poutanen}, J., \& {Johnson}, W.~N. 2000, \apj, 542, 703

\end{thebibliography}

\clearpage

\appendix

\section{Spectral extraction method}\label{app:extr}
We have developed a method to extract the averaged 
long term spectrum of a source. 
In this method, the spectrum is obtained as a weighted average of the source 
spectra of all observations where the source is in the field of view.
In particular, the averaged source  count rates in the i-th energy channel, 
$\bar{R_i}$,
and their error $\bar{\sigma_i}$,
are given by the following equations: 
\begin{equation}
\bar{R_i} = \frac{\sum_{j=0}^{N} r_j * w_j}{\sum_{j=0}^{N} w_j}\ , \ \ \ \
\bar{\sigma_i} = \sqrt{\frac{\sum_{j=0}^N w_j V_j}{N \sum_{j=0}^{N} w_j} }
\end{equation}
where $r_j$ is the source count rate in the j-th observation,
 $w_j$ is the weight used and the sums extend over all observations
which contain the source. 
Using the inverse of the  count rate variance $V_j$ as a weight,
the previous equations simplify to:
\begin{equation}
\bar{R_i} = \frac{\sum_{j=0}^{N} r_j \cdot 1/V_j}{\sum_{j=0}^{N} 1/V_j}\ , \ \ \
\bar{\sigma_i}  = \sqrt{\frac{1}{\sum_{j=0}^{N} 1/V_j} }
\label{eq:spec}
\end{equation}

However, the spectra entering in Eq. \ref{eq:spec} must be corrected
for off-axis count rate variation and for residual background contamination.
We explain below the way these corrections are implemented.

%
%
\subsection{Rate variation as a function of off-axis angle}

The detected count rates strongly vary with the position of the source in the
FOV; a source at the far edge of the partially coded FOV (PCFOV)
can experience a decrease in rate
of a factor $~2$ (depending also on energy) when compared to its
on-axis rate.\\

\begin{figure}[h!]
\epsscale{0.8}
\plotone{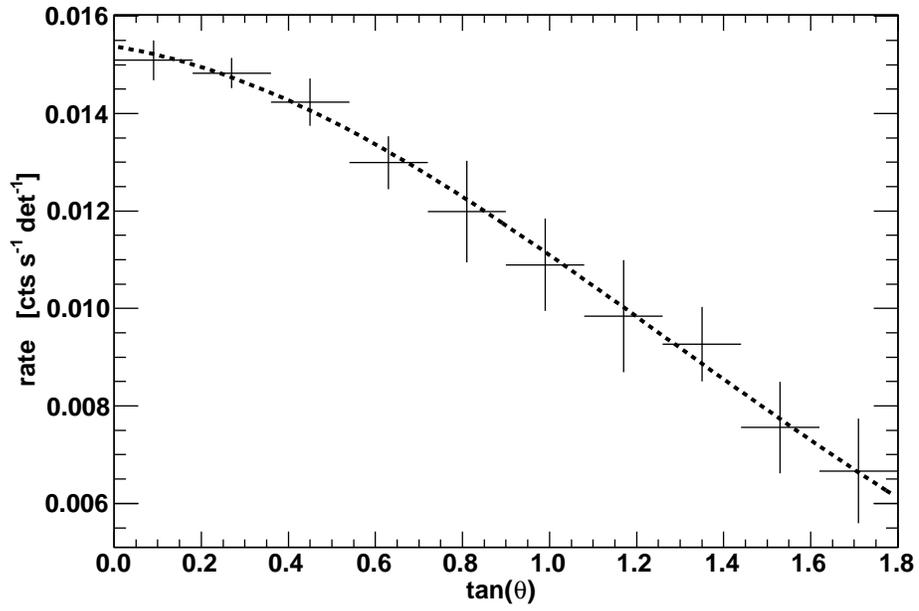}
\caption{Crab rates in the 14-22\,keV band as a function of the tangent
of the  off-axis angle. 
When the Crab is 50$^{\circ}$ off-axis the detected count rate is
$\sim$30\% lower than the on-axis count rate.
The solid line is a polynomial fit to the rates.}
	\label{fig:crab_rate}
\end{figure}

The standard Swift-BAT imaging software corrects for geometrical 
off-axis effects like
cosine and partial coding (vignetting) effects; it is only when the response matrix
is generated (tool {\it batdrmgen}) that other effects like detector 
thickness and effective area variation are  taken into account. \\
Since in equation \ref{eq:spec} we are averaging over spectra at
different positions in the FOV, we need to take into account 
the variations in the rates produced by the detector response.
In order to do so, we have analyzed a series of more than
1000 Crab observations.
For each
of our 6 energy channels we made a polynomial fit to the Crab rate as a 
function of the off-axis angle, and derived a set of corrective coefficients.
These coefficients are then used to correct the rates of each source spectrum
in order to transform  them to the  equivalent on-axis rates.\\
The variation of the Crab rates as a function of position in the FOV 
is reported in Fig. \ref{fig:crab_rate}.

%
%
%
\subsection{Residual background contamination}
In order to extract a source spectrum from survey data (in form of
Detector Plane Histograms, DPH) the user must first produce a mask
of weights (tool {\it batmaskwtimg}) for the source position and then
use this mask to extract the detected counts from the array (tool {\it batbinevt}).
The weights are chosen such  that the resulting spectrum is 
already background-subtracted. This is an implementation of the standard
mask weighting technique  called {\it balanced correlation}
\citep{fenimore78}. The automatic background subtraction works
as long as the noise in the array is flat and not correlated with the mask
pattern. These conditions are not always satisfied and a small background
contamination can arise.\\\\
The total background contamination for the case of the Crab is $<2$\%
when compared to the Crab on-axis rate in the 14-195\,keV band. Thus,
this contamination does not pose problems
for bright sources. However,   it becomes relevant 
for the spectral analysis of faint objects with intensities of $\sim$mCrab.
\\
In order to correct for this residual background contamination, we 
fit the {\it batclean} background model to each energy channel
in order to create a background prediction for each of them.
Convolving these background predictions with the mask of weights generated
for the source under analysis yields the residual background term which 
the mask weighting technique did not manage to suppress.

\subsection{Spectral fitting}

The final source rates in the $i$-th energy channel are computed as:
\begin{equation}
\bar{R_i} =\frac{\sum_{j=0}^{N} (r_j-b_j)\cdot K(E,\theta) \cdot 1/V_j}{\sum_{j=0}^{N} 1/V_j} 
\end{equation}

where $b_j$ is the residual background term, $K(E,\theta)$ is the parametrized
instrumental response as function of the energy channel and the off-axis angle
and $V_j$ is the rate variance. 
The weighted averaged spectrum is then input, together with a 
the BAT response matrix, to XSPEC 11.3.2 \citep{arnaud96} for spectral fitting.
\\\\
Finally, we checked that the averaged Crab spectrum, obtained with the 
above method, is consistent with the standard (BAT) Crab spectrum as detected
in each observation (photon index of 2.15 and normalization of 
10.15\,photons cm$^{-2}$ s$^{-1}$ at 1 keV
in the 15--200\,keV energy range).

%
%

\subsection{Notes on individual sources}\label{app:spec}
\begin{figure*}
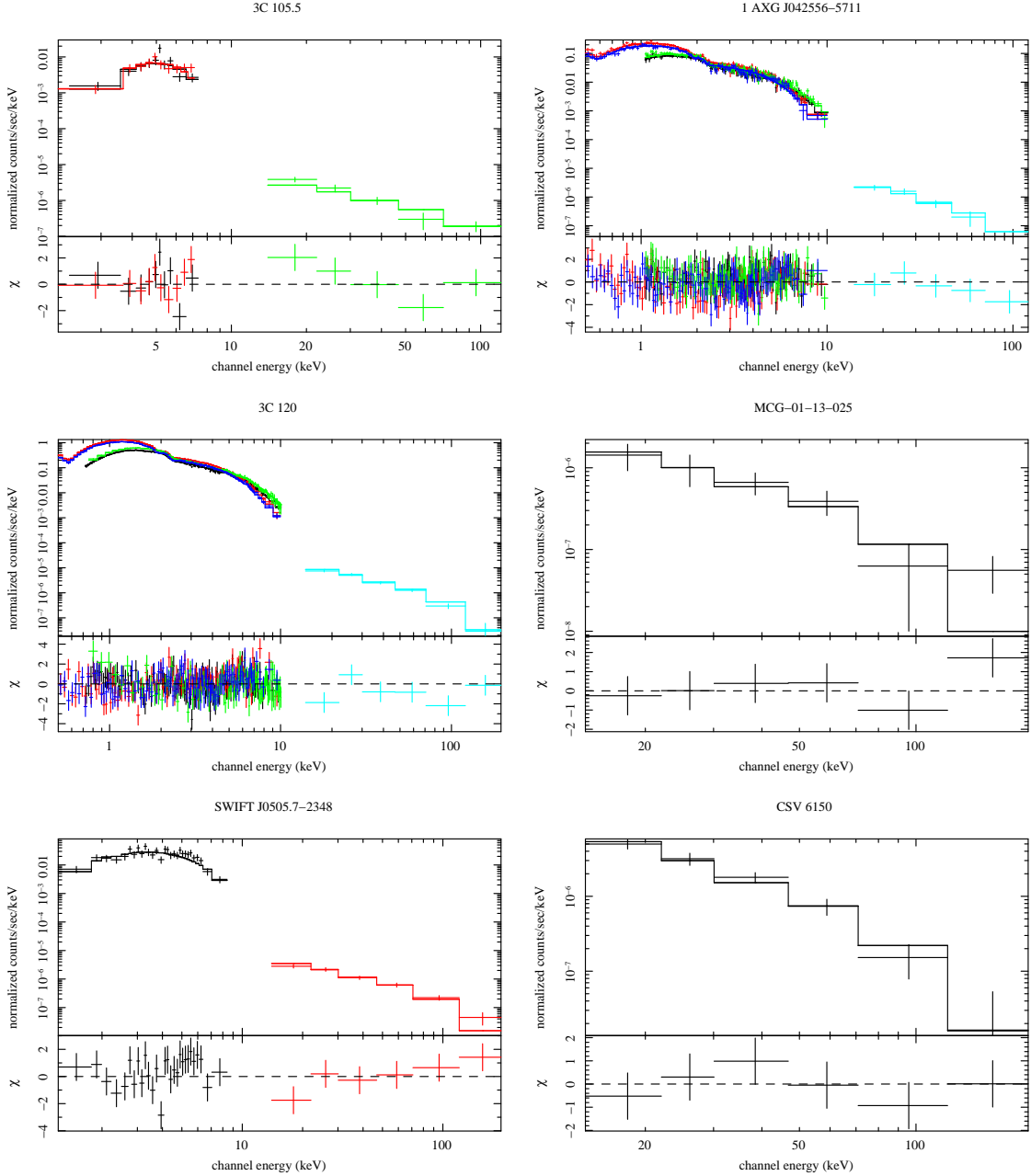
 
    \begin{center}
     \begin{tabular}{cc}
    \includegraphics[scale=0.3,angle=270]{f7a.ps} &
	   \includegraphics[scale=0.3,angle=270]{f7b.ps} \\
  \includegraphics[scale=0.3,angle=270]{f7c.ps} &
 \includegraphics[scale=0.3,angle=270]{f7d.ps} \\
  \includegraphics[scale=0.3,angle=270]{f7e.ps} &
  \includegraphics[scale=0.3,angle=270]{f7f.ps} \\
     \end{tabular}
    \end{center}
    \null\vspace{-7mm}
    \caption{Folded spectra and best fit models as described in the text. 
From left to right and up to
bottom the spectra are for: 
3C105.0, 1AXG J042556-5711,
3C120, MCG-01-13-025,
SWIFT J0505.7-2348 and CSV6150.
}
  \label{fig:spe1}
\end{figure*}

\begin{figure*}
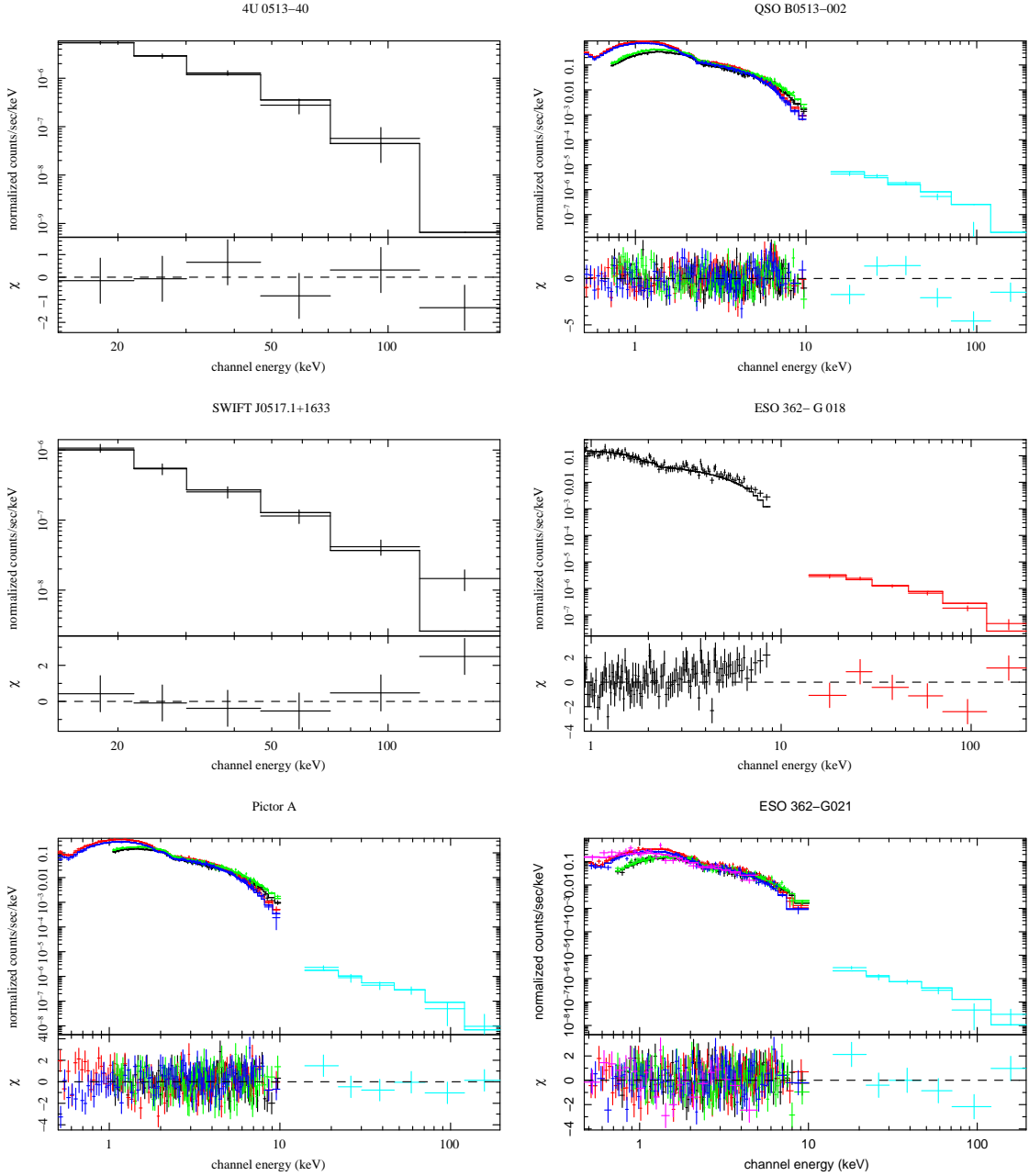
 
    \begin{center}
     \begin{tabular}{cc}
   \includegraphics[scale=0.3,angle=270]{f8a.ps} &
    \includegraphics[scale=0.3,angle=270]{f8b.ps} \\
	  \includegraphics[scale=0.3,angle=270]{f8c.ps} &
  \includegraphics[scale=0.3,angle=270]{f8d.ps}  \\
  \includegraphics[scale=0.3,angle=270]{f8e.ps} &
\includegraphics[scale=0.3,angle=270]{f8f.ps}  \\
  \end{tabular}
    \end{center}
    \null\vspace{-7mm}
    \caption{{\bf (for online version)} Folded spectra and best fit models as described in the text. 
From left to right and up to
bottom the spectra are for: 
4U 0513-40, QSO B0513-002,
SWIFT J0517.1+1633, ESO 362- G 018,
Pictor A and ESO 362- G 021.
}
  \label{fig:spe2}
\end{figure*}

\begin{figure*}
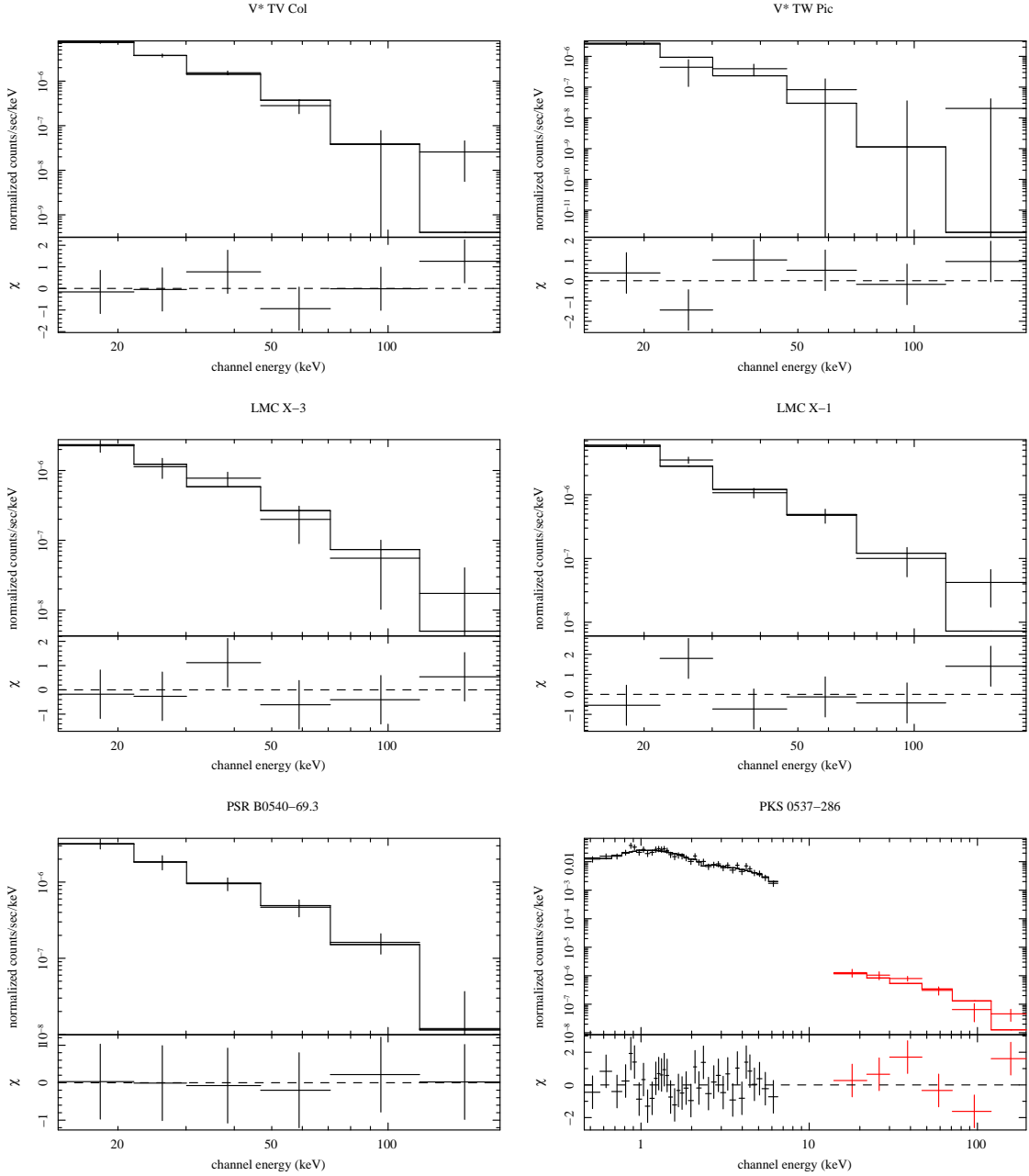
 
    \begin{center}
     \begin{tabular}{cc}
    \includegraphics[scale=0.3,angle=270]{f9a.ps} &
    \includegraphics[scale=0.3,angle=270]{f9b.ps} \\
    \includegraphics[scale=0.3,angle=270]{f9c.ps} &
    \includegraphics[scale=0.3,angle=270]{f9d.ps} \\
    \includegraphics[scale=0.3,angle=270]{f9e.ps} &
  \includegraphics[scale=0.3,angle=270]{f9f.ps} \\
     \end{tabular}
    \end{center}
    \null\vspace{-7mm}
    \caption{{\bf (for online version)} Folded spectra and best fit models as described in the text. 
From left to right and up to bottom the spectra are for: 
 TV Col,  TW Pic,
LMC X-3, LMC X-1,
PSR B0540-69.3 and PKS 0537-286.
}
  \label{fig:spe3}
\end{figure*}

\begin{figure*}
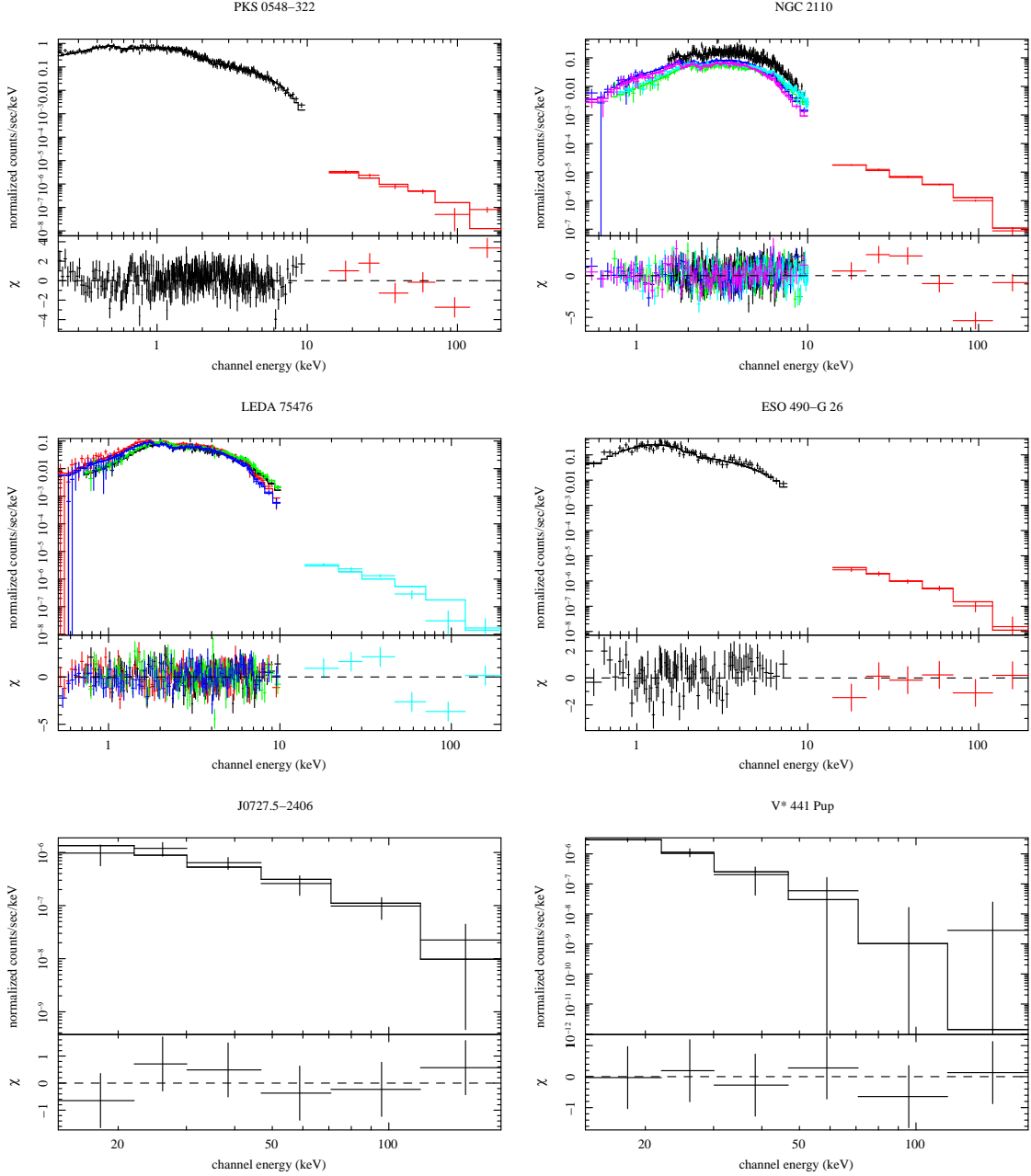
 
    \begin{center}
     \begin{tabular}{cc}
    \includegraphics[scale=0.3,angle=270]{f10a.ps} &
    \includegraphics[scale=0.3,angle=270]{f10b.ps} \\
 \includegraphics[scale=0.3,angle=270]{f10c.ps} &
 \includegraphics[scale=0.3,angle=270]{f10d.ps} \\
 \includegraphics[scale=0.3,angle=270]{f10e.ps} &
\includegraphics[scale=0.3,angle=270]{f10f.ps} \\

     \end{tabular}
    \end{center}
    \null\vspace{-7mm}
    \caption{{\bf (for online version)} 
Folded spectra and best fit models as described in the text. 
From left to right and up to bottom the spectra are for: 
PKS 0548-322, NGC 2110,
LEDA 75476, ESO 490- G 26,
J0727.5-2406 and  V441 Pup
}
  \label{fig:spe4}
\end{figure*}

\begin{figure*}
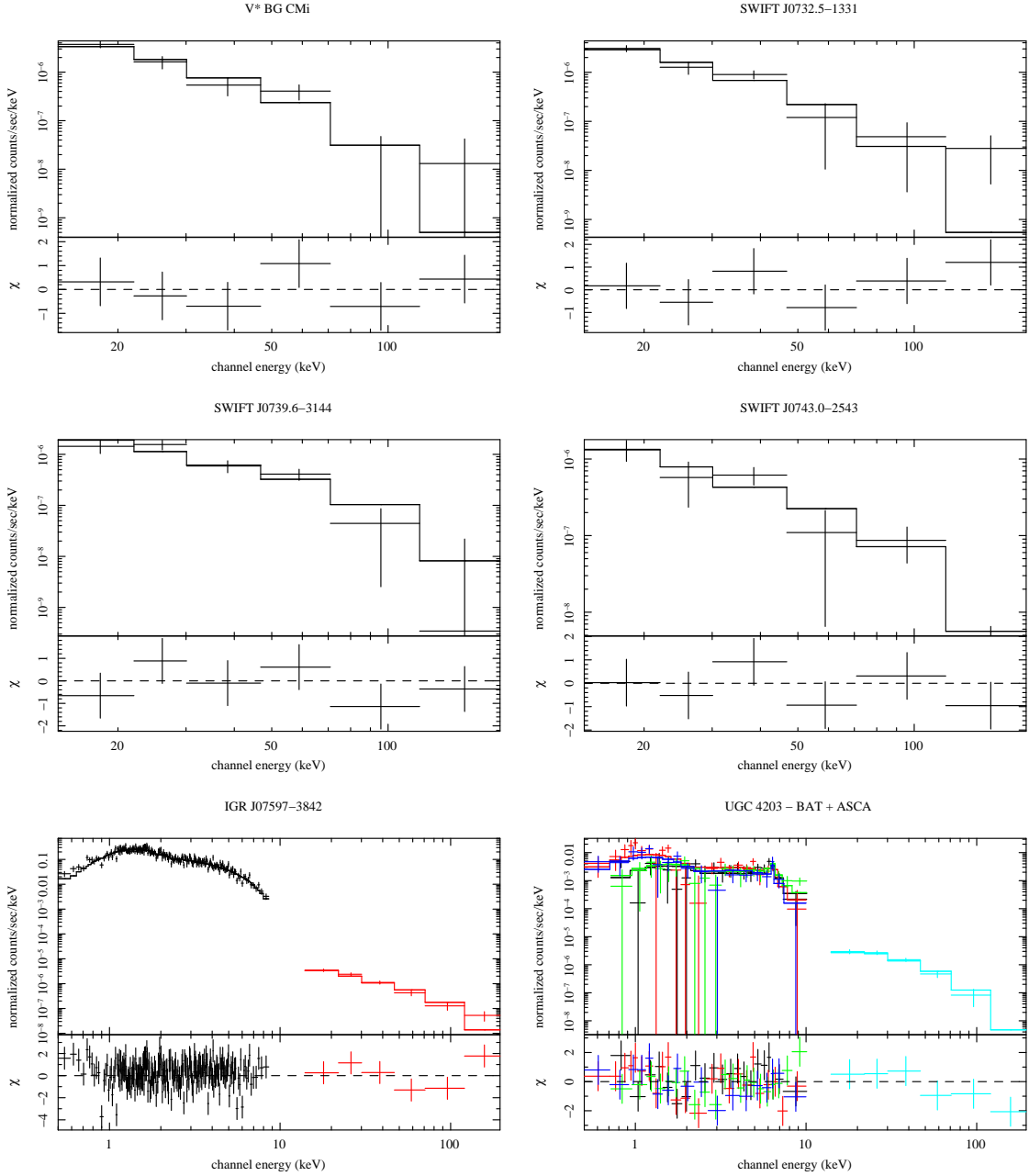
 
    \begin{center}
     \begin{tabular}{cc}
    \includegraphics[scale=0.3,angle=270]{f11a.ps} &
    \includegraphics[scale=0.3,angle=270]{f11b.ps} \\
   \includegraphics[scale=0.3,angle=270]{f11c.ps} &
   \includegraphics[scale=0.3,angle=270]{f11d.ps} \\
   \includegraphics[scale=0.3,angle=270]{f11e.ps} &
   \includegraphics[scale=0.3,angle=270]{f11f.ps} \\
     \end{tabular}
    \end{center}
    \null\vspace{-7mm}
    \caption{{\bf (for online version)} 
Folded spectra and best fit models as described in the text. 
From left to right and up to bottom the spectra are for: 
 BG CMi,  J0732.5-1331,
J0739.6-3144, J0743.0-2543,
IGR K07597-3842 and UGC 4203.
}
  \label{fig:spe5}
\end{figure*}

\begin{figure*}
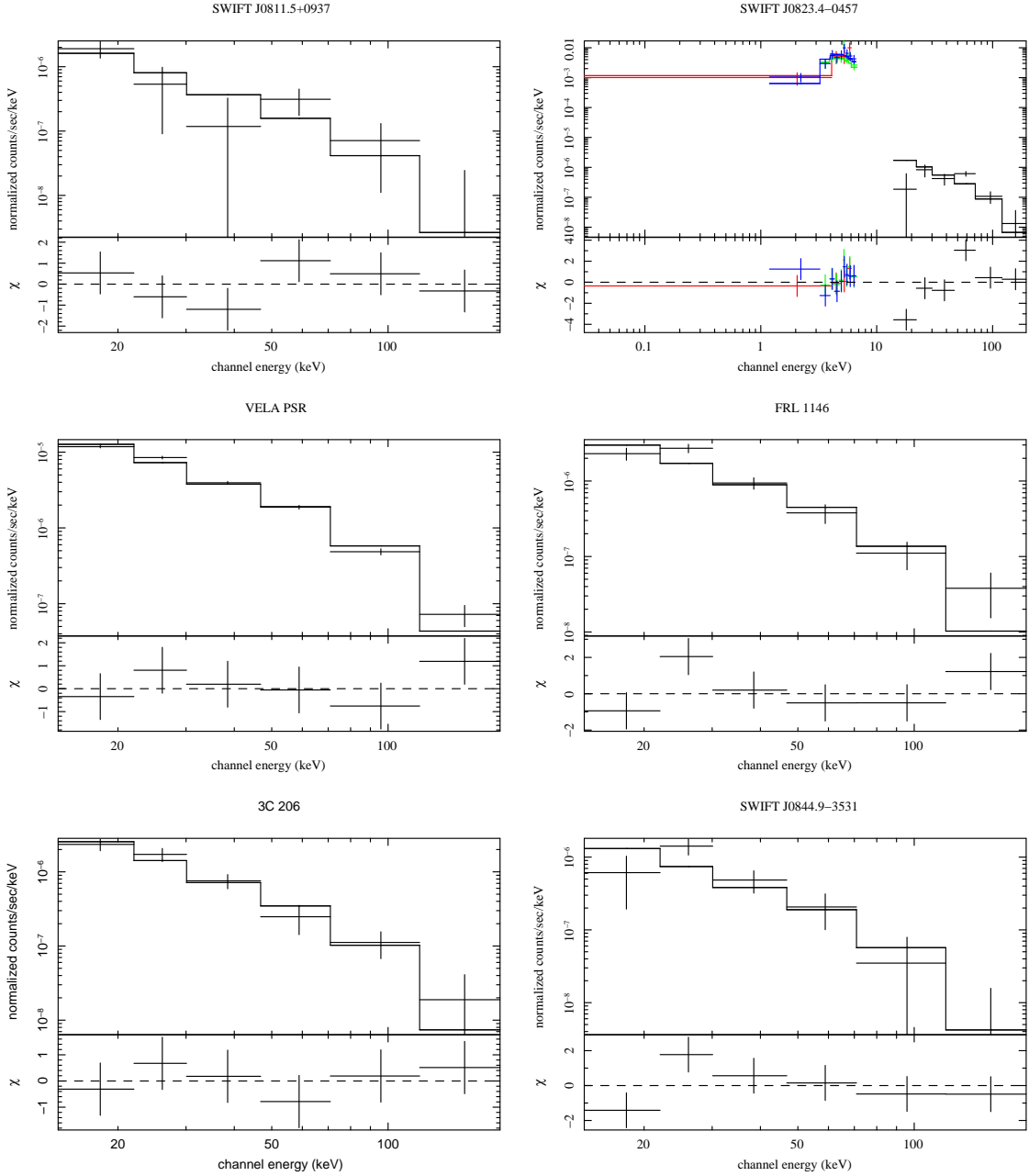
 
    \begin{center}
     \begin{tabular}{cc}
    \includegraphics[scale=0.3,angle=270]{f12a.ps} &
	 \includegraphics[scale=0.3,angle=270]{f12b.ps} \\
    \includegraphics[scale=0.3,angle=270]{f12c.ps} &
   \includegraphics[scale=0.3,angle=270]{f12d.ps} \\
    \includegraphics[scale=0.3,angle=270]{f12e.ps} &
 \includegraphics[scale=0.3,angle=270]{f12f.ps} \\
     \end{tabular}
    \end{center}
    \null\vspace{-7mm}
    \caption{{\bf (for online version)} 
Folded spectra and best fit models as described in the text. 
From left to right and up to bottom the spectra are for: 
J0811.5+0937, J0823.4-0457,
VELA PSR, FRL 1146,
3C 206 and J0844.9-3531.
}
  \label{fig:spe6}
\end{figure*}

\begin{figure*}
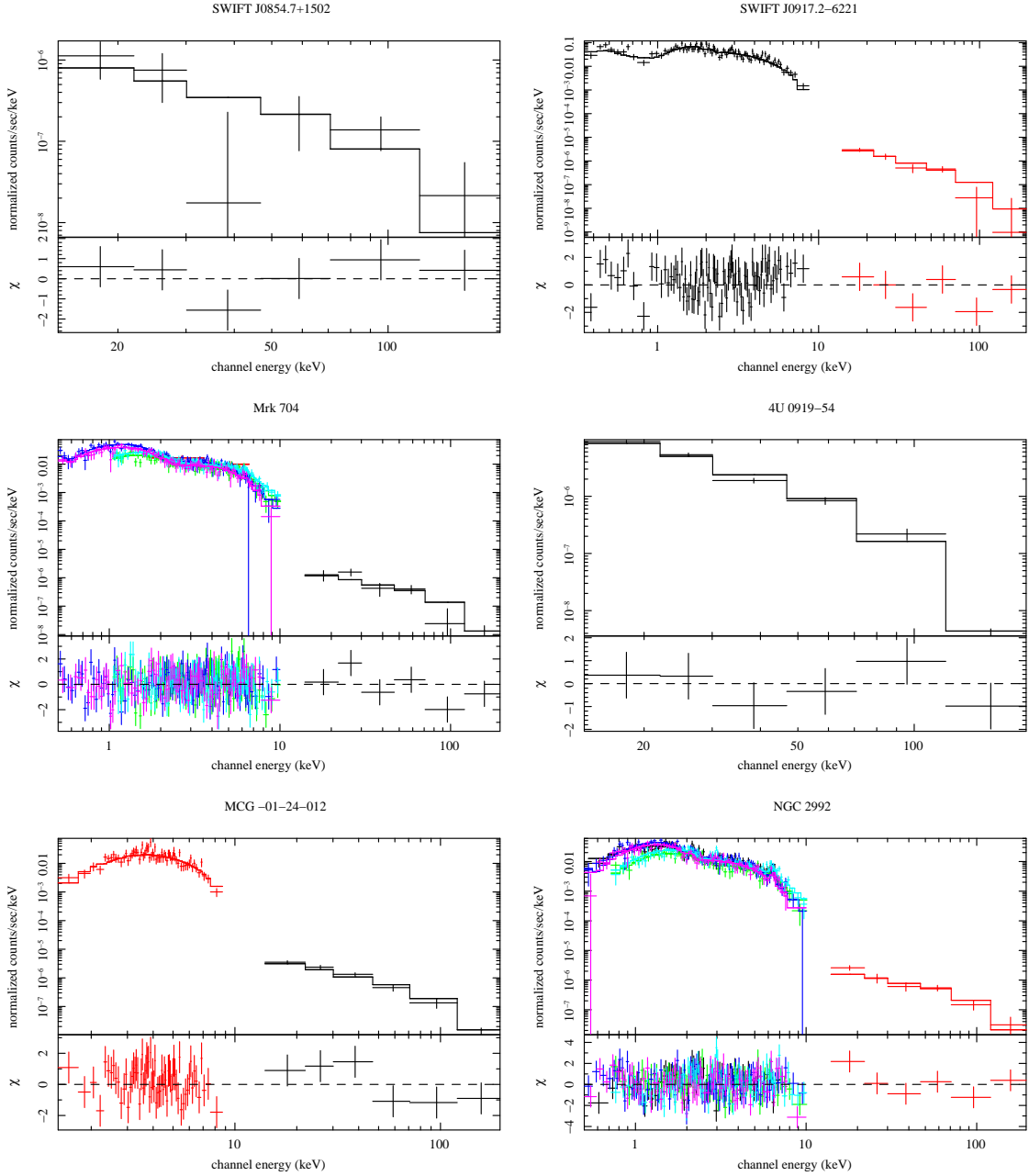
 
    \begin{center}
     \begin{tabular}{cc}
    \includegraphics[scale=0.3,angle=270]{f13a.ps} &
  \includegraphics[scale=0.3,angle=270]{f13b.ps} \\
 \includegraphics[scale=0.3,angle=270]{f13c.ps} &
 \includegraphics[scale=0.3,angle=270]{f13d.ps} \\
\includegraphics[scale=0.3,angle=270]{f13e.ps} &
  \includegraphics[scale=0.3,angle=270]{f13f.ps} \\
     \end{tabular}
    \end{center}
    \null\vspace{-7mm}
    \caption{{\bf (for online version)}
Folded spectra and best fit models as described in the text. 
From left to right and up to bottom the spectra are for: 
J0854.7+1502, J0917.2-6221,
Mrk 704, 4U 0919-54,
MCG -01-24-012 and NGC 2992.
}
  \label{fig:spe7}
\end{figure*}

We report a brief description of the source spectra for all new, or
interesting, sources found in this analysis. All quoted errors are 90\%.
The spectra of all the sources are reported in Figures  
\ref{fig:spe1}, \ref{fig:spe2}, \ref{fig:spe3},  \ref{fig:spe4},  
\ref{fig:spe5}, \ref{fig:spe6} , \ref{fig:spe7} and \ref{fig:spe7} 
(available online).
\\\\
{\bf 3C 105.0} is a Sy2 galaxy. The BAT and XRT  data 
can be fitted by an absorbed  power law model
with photon index of 1.65$\pm 0.13$ and hydrogen column density 
of $29.4^{+5.7}_{-4.8} \times 10^{22}$  atoms cm$^{-2}$.
Given its absorption and its luminosity (4.45$\times 10^{44}$\,erg s$^{-1}$),
3C 105.0 is a highly-absorbed highly-luminous QSO.
\\\\
{\bf 1 AXG J042556-5711} (also known as 1H 0419-577LB 1727, 
1ES 0425-573 and IRAS F04250-5718) is a radio-quiet Seyfert galaxy
which has been observed over recent years by ASCA, ROSAT, BeppoSAX  and recently
also by XTE \citep{revnivtsev06}. The ASCA and BAT data are well fit by an 
unabsorbed cut-off power-law model 
with photon index of 1.54$\pm0.028$ and cut-off at 73$^{+46.2}_{-24.1}$\,keV.
\\\\
{\bf 3C 120} is a Sy1 galaxy. This source was observed by ASCA.
The best fit to ASCA and BAT data is an absorbed power law model
with absorption consistent with the galactic one and photon index
$1.80^{+0.04}_{-0.04}$ and a black body component with a 
temperature of $0.27^{+0.026}_{-0.025}$\,keV.
\\\\
{\bf MCG  -01-13-025} is a Sy1.2 (in NED, but Sy1 in SIMBAD) 
galaxy detected in soft X-rays by ROSAT \citep{voges99}.
The BAT spectrum is consistent with a power law with photon index of 
$ 1.6^{+0.48}_{-0.47}$ and it extends up to 200\,keV.
\\\\
{\bf SWIFT J0505.7-2348}, also know as XSS J05054-2348 
\citep{revnivtsev06}  is a Sy2 galaxy.
When combining both XRT and BAT data for this source we get an intrinsic,
 rest frame, absorption of 
4.8$^{+0.9}_{-0.7} \times 10^{22}$\,atoms cm$^{-2}$ and
a photon index of 1.77$^{+0.08}_{-0.07}$.
\\\\
{\bf CSV 6150}, also known as IRAS 05078+1626, is cataloged as Sy1 in SIMBAD
and as Sy1.5 in NED. The BAT spectrum can be fitted
with a power law with photon index of  1.94$^{+0.25}_{-0.23}$.
The source flux in the 14--170\,keV band is 
6.3$^{+0.7}_{-4.0}\times 10^{-11}$\,erg cm$^{-2}$ s$^{-1}$ while
the luminosity is 4.4$^{+0.6}_{-2.2} \times 10^{43}$ erg s$^{-1}$.
\\\\
{\bf 4U 0513-40} is a low mass X-ray binary detected in X-rays 
by EXOSAT \citep{giommi91}. The BAT spectrum
can be fit by a  bremsstrahlung model with temperature of 
29.7$^{+7.5}_{-5.8}$\,keV. 
\\\\
{\bf QSO B0513-002} is a Sy1 galaxy. The BAT and ASCA spectra 
can be fitted by an absorbed  power-law model and a black body component.
The required absorption is in agreement with the galactic one. The photon
index and the plasma temperature are respectively 
$1.83^{+0.02}_{-0.016}$ and $0.27^{+0.02}_{-0.02}$\,keV. We also detect
a an iron line whose equivalent width is 90.8$^{+66.2}_{-76.7}$\,eV.
\\\\
{\bf SWIFT J0517.1+1633} is a new hard X-ray source \citep{ajello07a}. 
The BAT spectrum is
best fit by a power law model with photon index of 2.0$^{+0.23}_{-0.26}$.
\\\\
{\bf ESO 362- G 018} is a Sy1 galaxy detected at hard X-ray by BAT 
\citep{tueller05b}. The BAT and XRT data are best
fit by an absorbed power-law model with photon index of 1.50$^{+0.03}_{-0.02}$ 
and absorption consistent with the galactic value.
\\\\
{\bf Pictor A} is a radio-loud Sy1 galaxy initially detected in X-ray by 
the EINSTEIN observatory \citep{elvis92}. 
The best fit to ASCA and BAT data is an absorbed power-law model
with photon index of  $1.8^{+0.02}_{-0.02}$ and intrinsic
absorption of $1.14^{+0.01}_{-0.05}\times 10^{21}$atoms cm$^{-2}$
slightly in excess of the galactic one ($~4\times 10^{20}$\,atoms cm$^{-2}$).
\\\\
{\bf ESO 362-G021} is a BL Lac object. ASCA 
and XRT data are available for this source.
The best fit to ASCA, BAT and XRT data is an absorbed power law with photon
index of  $1.72^{+0.04}_{-0.04} $ and intrinsic column density of 
$ 0.14^{+0.02}_{-0.02}\times 10^{22}$\,atoms cm$^{-2}$.
\\\\
{\bf TV Col} is a DQ Her type  cataclysmic 
variable already detected at soft and 
hard X-rays. A power law fit to the BAT spectrum does not yield acceptable
results; instead a bremsstrahlung model with a plasma temperature 
of $28.2^{+4.6}_{-3.8}$\,keV fits the data well.
\\\\
{\bf  TW PIC} is a cataclysmic variable of the DQ Her type \citep{norton00}.
The BAT spectrum is best fitted by a bremsstrahlung model with 
a plasma temperature of $13.5^{+10.6}_{-5.6}$\,keV. The flux  of 
5.5$\times 10^{-12}$erg cm$^{-2}$ s$^{-1}$ in the 20-40\,keV band 
is a factor 2 lower than the one reported in a recent
INTEGRAL measurement \citep{gotz06} suggesting variability.
\\\\
{\bf LMC X-3} is high mass X-ray binary (HXB). 
The BAT spectrum is consistent with 
a power law whose photon index is 2.0$^{+0.4}_{-0.3}$.
\\\\
{\bf LMC X-1} is a well known black hole candidate. It is detected up to
200\,keV with a steep photon index of 2.3$^{+0.22}_{-0.20}$. The flux is a 
factor 2 lower than the one measured by INTEGRAL \citep{gotz06}, suggesting
variability.
\\\\
{\bf PSR B0540-69.3} is a young rotation-powered pulsar recently detected up to
60\,keV also by INTEGRAL \citep{gotz06,slowikowska06}. 
The Pulsar is detected in BAT up
to 200\,keV and it is spectrum can be modeled as a power law with photon index
of 1.85$^{+0.28}_{-0.26}$.
\\\\
{\bf PKS 0537-286} at z=3.1 is one of the most luminous high-redshift quasar.
Recognized first as a radio source \citep{bolton75} it was 
discovered in X-rays by the Einstein observatory \citep{zamorani81}
and then studied by ROSAT, ASCA and lately by XMM. The BAT detection in hard 
X-rays is the first to date, however there is a claim that PKS 0537-286 be
the MeV counterpart of the EGRET source  3EG J0531-2940  \citep{sowards04}.
A joint spectral fit to XRT and BAT data reveals an exceptionally hard 
spectral slope of $1.35^{+0.06}_{-0.08}$.
\\\\
{\bf PKS 0548-322} is a well known blazar already detected  in hard X-rays 
(see for example Donato et al., 2005).  A joint spectrum of XRT and BAT 
data with an absorbed power law model yields a photon index of 
$1.8^{+0.03}_{-0.03}$ and an intrinsic absorption   of 
2.57$^{+0.6}_{-0.5}\times 10^{20}$ atoms cm$^{-2}$.
\\\\
{\bf NGC 2110} is a well known Sy2 galaxy. The BAT, ASCA and XRT  data
can be fit by an absorbed power law model (photon index of
$1.62^{+0.01}_{-0.01}$ and intrinsic hydrogen column density of 
$4.0^{+0.13}_{-0.07} \times 10^{22}$ atoms cm$^{-2}$) with a soft excess
which could be described as black body component with temperature of
0.47$^{+0.02}_{-0.02}$. 
We also detected an unresolved
Fe K$_{\alpha}$ of equivalent width of 118$^{+42}_{-53}$ eV.
\\\\
{\bf LEDA 75476}, also known as 3A 0557-383, EXO 055620-3820.2 and CTS B31.01,
is a Sy1 galaxy. The BAT spectrum is consistent with a power law
model with photon index of $2.0\pm0.4$.
The ASCA and BAT are well fit by an absorbed power law model with 
photon index of 1.74$^{+0.02}_{-0.03}$ and intrinsic 
absorbing column density of 
$2.2^{+0.11}_{-0.13} \times 10^{22}$atoms cm$^{-2}$. A clear excess
below 2\,keV is detected in the ASCA data and this can be modeled as 
a black body component with a temperature of $0.28^{+0.08}_{-0.05}$\,keV.
A Fe K$_{\alpha}$ line is also required by the fit (F-test yielding 
a probability of the line being spurious of 10$^{-8}$) and its equivalent
width is 0.132\,keV. The reduced $\chi^2$ of the overall fit is 1.1.
\\\\
{\bf ESO 490-G 26} is a Sy1.2 galaxy.
The joint XRT-BAT spectrum can be described as a power law 
with photon index of 1.90$^{+0.05}_{-0.04}$ 
and an intrinsic, in addition to the 
galactic, absorption of 2.7$^{+0.05}_{-0.05} \times 10^{21}$\,atoms cm$^{-2}$.
The flux and the luminosity in the 14--170\,keV band are 
3.6$^{+1.1}_{-1.3}\times 10^{-11}$\,erg cm$^{-2}$ s$^{-1}$ and 
4.7$^{+1.2}_{-3.6}\times 10^{43}$\,erg s$^{-1}$. 
\\\\
{\bf SWIFT J0727.5-2406} has a spectrum consistent with a power law model with
photon index of $1.53\pm0.54$. As already noted by \cite{rau07} this
BXS source is likely associated with the nearby ROSAT source
1RXS J072720.8-240629 and with the radio object NVSS J072721-240632.
\\\\
{\bf  V441 Pup} is a high mass X-ray binary where the companion was optically
identified as a Be star. The BAT spectrum is very steep and it can either
be fitted by a power law with a photon index of 4.5$\pm1.5$
or by a  bremsstrahlung model with a plasma temperature of 
$12.4^{+13.6}_{-5.6}$\,keV.
\\\\
{\bf  BG CMi} is a well known intermediate polar. The BAT spectrum is
consistent with a bremsstrahlung model with a plasma temperature of
 31.3$^{+41.2}_{-14.2}$\,keV.
\\\\
{\bf SWIFT J0732.5-1331} was detected for the first time 
by BAT in hard X-rays \citep{ajello06a}. 
It was then identified as a new intermediate polar
(Wheatley et al. 2006 and references therein). The BAT spectrum
is consistent with a bremsstrahlung model with a plasma temperature 
of $33.2^{+50.1}_{-14.2}$\,keV.
\\\\
{\bf SWIFT J0739.6-3144} is a newly discovered hard X-ray source \citep{ajello07a},
recently identified as a Sy2 galaxy \citep{rau07}.
A simple power law fit to the BAT spectrum yields a photon index of
$1.77^{+0.51}_{-0.43}$ . 
We also estimated the  lower limit on the absorbing column density 
considering the undetection
by ROSAT; this limit is $\sim2 \times 10^{22}$  atoms cm$^{-2}$.
The flux and the luminosity in the 14--170\,keV band are 
2.3$^{+1.1}_{-1.8}\times 10^{-11}$\,erg cm$^{-2}$ s$^{-1}$ and 
3.2$^{+1.6}_{-1.9}\times 10^{43}$\,erg s$^{-1}$. 
\\\\
{\bf SWIFT J0743.0-2543} is a newly discovered hard X-ray source \citep{ajello07a}.
The BAT spectrum is consistent with a power-law model with
photon index of 1.78$^{+0.69}_{-0.56}$.
As noted in \cite{rau07} this BXS source is likely to be associated
with the ROSAT source 1RXS J074315.6-254545 and the galaxy  LEDA 86073.
\\\\
{\bf IGR J07597-3842} is a source first detected by INTEGRAL 
in the VELA region 
\citep{denhartog04}. It was identified as being a Sy1.2 \citep{masetti06b}.
This source was also observed by XRT and when jointly fitting XRT and BAT data
we get that the best fit is an absorbed power law with photon index 
of $1.8^{+0.08}_{-0.07}$
and column density of $5.8^{+0.5}_{-0.5} \times 10^{21}$atoms cm$^{-2}$ 
consistent with the galactic foreground absorption. The source is thus
unabsorbed.
The flux and the luminosity in the 14--170\,keV band are 
4.2$^{+0.6}_{-2.4}\times 10^{-11}$\,erg cm$^{-2}$ s$^{-1}$ and 
15.9$^{+1.5}_{-14.8}\times 10^{43}$\,erg s$^{-1}$. 
\\\\
{\bf UGC 4203} is a Sy2 galaxy. 
As already noted in \cite{matt03}, this source shows transitions
between a reflection-dominated and a transmission-dominated spectrum.
The ASCA and BAT data can be successfully fit by a reflection model
\citep[PEXRAV,][]{magdziarz95}
 with photon index of 1.68$\pm0.1$ and a reflection
normalization of 65.2$^{43.2}_{-23.07}$  and a prominent iron line 
with equivalent width of 0.7$^{+1.1}_{-0.6}$\,keV.
A soft excess at energies $<1$\,keV can be modeled as a black body component with a temperature of $0.3^{+0.08}_{-0.05}$\,keV.
\\
XRT data are also available for this source. However, the XRT spectrum
has a lower quality than the ASCA one. In the XRT observation, the source
is found in a transmission-dominated state; the best fit model is an absorbed
reflection model (the reflection component is required by the BAT spectrum)
with hydrogen column density
of N$_H$=12.5$^{+5.0}_{-3.7} \times 10^{22}$\,atoms cm$^{-2}$, photon index
of 2.0$^{+0.25}_{-0.25}$ and reflection normalization of 2.12$^{+2.5}_{-1.5}$.
\\\\
{\bf SWIFT J0811.5+0937}  is a new BXS source detected in \cite{ajello07a}. 
The BAT spectrum is consistent with a power law with photon index of 
$2.2^{+2.1}_{-0.9}$. \cite{rau07} identified RX J081132.4+093403 as 
possible counterpart. Optical spectroscopy revealed that this source
is a candidate X-ray Bright Optically Normal Galaxy (XBONG). 
If we extrapolate the BAT power law to the ROSAT-PSPC energy band 
(0.1--2.4\,keV) we get no indication of intrinsic absorption.
\\\\
{\bf SWIFT J0823.4-0457} is a source detected for the first time 
in hard X-rays by BAT and associated, during an XRT follow-up,
with the galaxy FAIRALL 0272 \citep{ajello07a}. An optical 
follow-up showed that the source is a  Sy2 \citep{masetti06}.
XRT and BAT data are best fit by a highly absorbed power law. The photon 
index is $1.84^{+0.28}_{-0.22}$ and the absorbing column density is 
$19.3^{+6.8}_{-5.4} \times 10^{22}$\,atoms cm$^{-2}$.
\\\\
{\bf Vela PSR} has a spectrum consistent with a power law whose photon index
is 1.88$\pm 0.2$.
\\\\
{\bf FRL 1146} is a Sy1 galaxy deteced in hard X-ray by INTEGRAL 
\citep{bird06}.
The BAT spectrum is characterized by a power law with photon index of 
$1.88^{+0.37}_{-0.31}$
extending up to 200\,keV. The 14--170\,keV flux and luminosity of 
3.3$^{+0.8}_{-0.7}\times 10^{-11}$\,erg cm$^{-2}$ s$^{-1}$ and 
7.2$^{+1.6}_{-1.4}\times 10^{43}$\,erg s$^{-1}$ 
are in agreement with the  INTEGRAL measurement.
FRL 1146 was also detected in the ROSAT all-sky survey at 
12\,count s$^{-1}$, 
considering the extrapolation of the BAT power law to the ROSAT band yields 
$\sim$8\,count s$^{-1}$ so it is very likely that the source in unabsorbed. 
\\\\
{\bf 3C 206} is a narrow line, radio loud, 
QSO detected for the first time in hard X-rays
($>$20\,keV). It was detected by Lawson \& Turner (1997) 
using GINGA in the 2-10\,keV
The BAT spectrum is consistent with a pure
power-law model with photon index of $1.95^{+0.43}_{-0.39}$.
3C 206 was detected by the ROSAT PSPC with 0.37\,ct s$^{-1}$ 
during the all-sky survey
\citep{voges99}; if we use the BAT power law spectrum and we 
extrapolate it to the
0.1-2.4\,keV band, we get that no additional absorption (with respect 
to the galactic one) is required to match the observed ROSAT count rate.
\\\\
{\bf SWIFT J0844.9-3531} is a new hard X-ray source detected in \cite{ajello07a}.
The BAT spectrum is consistent with a power law model with photon
index $1.91^{+0.46}_{-0.68}$.
The flux in the 14--170\,keV band is
1.7$^{+1.1}_{-0.6}\times 10^{-11}$\,erg cm$^{-2}$ s$^{-1}$. \cite{rau07}
noted that this BXS source might likely be associated with the ROSAT source
1RXS J084521.7-353048.
\\\\
{\bf SWIFT J0854.7+1502} is a new hard X-ray source detected in \cite{ajello07a}
and identified in \cite{rau07} as a Sy2 galaxy. 
It has a  flat spectrum which can be modeled as a power law with
photon index $1.41^{+0.7}_{-0.9}$.
A lower limit on the absorbing column density
of 5$\times 10^{21}$\,atoms cm$^{-2}$ can be derived by the non-detection
of this source in the ROSAT all-sky survey.
\\\\
{\bf SWIFT J0917.2-6221} is a new hard X-ray source. We analyzed a 7 ks
XRT observation of this source. The XRT and BAT data are well fit by an 
absorbed power law model whose photon index is $1.87^{+0.07}_{-0.04}$
and absorbing column density of 
1.33$^{+0.18}_{-0.10} \times 10^{22}$\,atoms cm$^{-2}$.
A clear excess is present at energies $<$1\,keV and this can be well described
as a black body component peaking at 0.14\,keV.
The flux and the luminosity in the 14--170\,keV band are 
2.6$^{+0.8}_{-0.8}\times 10^{-11}$\,erg cm$^{-2}$ s$^{-1}$ and 
20.0$^{+6.0}_{-5.0}\times 10^{43}$\,erg s$^{-1}$. 
\\\\
{\bf Mrk 0704}, or SWIFT 0918.5+1618,
is another source found thanks to our algorithm \citep{ajello07a}.
During an XRT follow-up, the galaxy Mrk 704 was found as the BAT counterpart.
Mrk 704 was previously detected in soft X-rays by ROSAT \citep{schwope00}.
In a recent optical follow-up, the galaxy was found to be a Sy1 \citep{masetti06}.
We have analyzed ASCA, XRT and BAT data for this source.
The best fit to the three datasets is a 
a partial covering model where the covering fraction
is 0.5 and the powerlaw photon index is $ 1.36^{+0.10}_{-0.07}$.
The source is highly
absorbed with a column density of 1.5$^{+0.6}_{-0.3} \times 10^{23}$  atoms cm$^{-2}$.
We also detected an iron line whose equivalent width is 160\,eV.
\\\\
{\bf 4U 0919-54}, detected at very high significance, is 
a LMXB also known to produce X-ray bursts \citep{jonker01}. Its spectrum
is characterized by a  steep photon index of $2.35 \pm 0.25$, alternatively
a bremsstrahlung model with a plasma temperature of 45.11$^{+26.13}_{-9.80}$\,keV
 yields a  better $\chi^2$.
\\\\
{\bf MCG -01-24-012} is a Sy2 galaxy already 
detected in hard X-rays by Beppo-SAX \citep{malizia02}.
When fitting both XRT and BAT data we get that the spectrum is consistent
with an absorbed power law whose   photon index is $ 1.7^{+0.08}_{-0.07}$  
and intrinsic absorption  is $6.5^{+0.8}_{-0.7}\times 10^{22}$\,atoms cm$^{-2}$.
\\\\
{\bf NGC 2992} is a Sy 1.9. The best fit for combined
XRT, ASCA and BAT data is an absorbed power law with photon 
index of $1.24^{+0.06}_{-0.05}$ 
and intrinsic hydrogen column density of $0.17^{-0.03}_{-0.03} \times 10^{22}$\,atoms cm$^{-2}$. 
We also detected the presence of an unresolved Fe K$_{\alpha}$
line whose equivalent width is $0.52^{+1.0}_{-0.1}$\,keV in agreement 
with an old Beppo-Sax measurement \citep{gilli01} where the reported
column density is 1$\times 10^{22}$atoms cm$^{-2}$.
\\\\
{\bf ESO 434-G 040} is a known Sy2 galaxy recently detected in hard X-rays 
also by INTEGRAL \citep{bird06}. A joint fit to ASCA, XRT and BAT data with an
absorbed power law model yields a photon index of $1.77^{+0.006}_{-0.07}$ and
a column density of $1.5^{+0.026}_{-0.09} \times 10^{22}$\,atoms cm$^{-2}$. 
A clear excess below 2\,keV can be modeled as a black body component with a 
temperature of $0.13^{+0.011}_{-0.016}$. An iron K$_{\alpha}$ line,
with an EQW=$85.5^{+27}_{-33}$, is also detected. The probability
of the line being spurious is $\sim 10^{-14}$.
\\\\
{\bf 3C 227} is a Sy1 galaxy and also a Radio galaxy.
The BAT spectrum 
is consistent with a power law model of  photon index $1.96^{+0.44}_{-0.58}$.
This source was detected at a level of 0.016\,ct s$^{-1}$ in a 11\,ks long
ROSAT-PSPC observation (0.1--2.4\,keV) \citep{crawford95}. 
In order to match the ROSAT observed count rates, the extrapolation
of the BAT power law to the 0.1--2.4\,keV band requires an absorbing column
density of at least $5\times 10^{21}$\,atoms cm$^{-2}$.
A recent Chandra observation confirms that 3C 227 is indeed an absorbed 
Sy1. However the joint Chandra-BAT spectrum is complex. 
Our best fit model is the sum of an absorbed power-law model 
and of a reflection component (both having the same photon index of
2.11$^{+0.14}_{-0.24}$). The absorbing column density is 
N$_H$=3.6$^{+1.5}_{-1.4} \times 10^{22}$\,atoms cm$^{-2}$.
The reflection component seems to be large R$>1$ which is at odds
with the absence of the iron K$_{\alpha}$ line. This source certainly
deserves further investigations.
\\\\
{\bf NGC 3081} is mis-catalogged in SIMBAD as Sy1 galaxy. In fact 
the available 6dF spectrum shows clearly that this object is a Sy2 object.
We have analized BeppoSax-MECS and ASCA data for this source.
The best fit is a
sum of a black body component, peaking at 
$0.58^{+0.15}_{-0.11}$\,keV, an absorbed power law
with column density of $60^{+3.1}_{-3.1} \times 10^{22}$\,atoms cm$^{-2}$ and
photon index $1.9^{+0.02}_{-0.04}$
and an iron line of equivalent width  of 241$^{+184}_{-131}$\,eV.


%
%
\begin{deluxetable}{lccccccccl}
\rotate
\tabletypesize{\scriptsize}
\tablewidth{0pt}
\tablecaption{Spectral parameters\label{tab:spec}}
\tablehead{
\colhead{NAME}           & \colhead{RA}     & \colhead{DEC}     & \colhead{Type}   &\colhead{$\Gamma$/E[kT]\tablenotemark{a}}   & \colhead{N$_H$}   &
\colhead{MODEL}          & \colhead{INSTR.\tablenotemark{b}}             \\
\colhead{} & \colhead{\scriptsize (J2000)}  & \colhead{\scriptsize (J2000)}    &
\colhead{} & \colhead{}  & 
\colhead{\scriptsize ($10^{22}$atoms cm$^{-2}$) }

}
\startdata 
%

3C 105.0    &61.9178 &  3.6517  & Sy2
&$1.66^{+0.13}_{-0.13}$  &$29.4^{+5.7}_{-4.8}$ &wabs*pow &B, X\\

1AXG J042556-5711  &66.6021 &-57.1775 &Sy1 & $1.54^{+0.028}_{-0.027}$ & 0 & pow &  B, A \\

3C 120        & 68.2982 &  5.3374  & Sy1 &$ 1.80^{+0.04}_{-0.04}$/$0.27^{+0.026}_{-0.025}$  &  0 & wabs*pow+bb & B, A \\

 MCG -01-13-025 & 72.9205 &-3.8240 & Sy1.2 &
$ 1.6^{+0.48}_{-0.47}$  & $ <0.02$ (1) & pow & B \\

 SWIFT J0505.7-2348 & 76.4674 & -23.8666 & Sy2 &
$ 1.77^{+0.08}_{-0.07}$ &$ 4.8^{+0.9}_{-0.7}$ 
& wabs*pow & B, X\\

CSV 6150    & 77.7224 & 16.5265 & Sy1.5  & 1.94$^{+0.25}_{-0.23}$ 
& \nodata & pow & B \\

4U 0513-40      & 78.5146 & -40.0558 & LXB      & 29.7$^{+7.5}_{-5.8}$   
& \nodata &  brem &B\\

QSO B0513-002  & 79.0096 & -0.1332 & Sy1
&$1.83^{+0.02}_{-0.016}$/$0.27^{+0.02}_{-0.02}$ &$<0.01$ & wabs*pow+bb & B, A \\

SWIFT  J0517.1+1633 & 79.2839 & 16.5605 & \nodata
& 2.0$^{+0.23}_{-0.26}$  & \nodata & pow & B \\

 ESO 362- G 018  & 79.8844 & -32.6720  & Sy1.5
& $1.5^{+0.03}_{-0.02}$  & $<$0.01 & wabs*pow & B,X \\

 Pictor A   &  79.9460 &  -45.7557   &  Sy1
&$1.8^{+0.015}_{-0.014}$ &$0.12^{+0.007}_{-0.02}$ &wabs*pow &B, A\\

ESO 362-G021    &80.6581  &  -36.4233 & BL Lac 
   &$1.7^{+0.037}_{-0.036} $ &$ 0.1^{+0.0197}_{-0.0187}$ & wabs*pow & B, A, X\\

V* TV Col & 82.3541 & -32.7965  & CV-DQ* & $24.9^{+4.6}_{-3.8}$ 
& \nodata & bremss & B \\

V* TW Pic      & 83.6470 & -58.0200 & CV &  
13.5$^{+10.6}_{-5.6}$ & \nodata & bremss & B \\

LMC X-3         & 84.7717 & -64.1148 & HXB  & 2.0$^{+0.4}_{-0.3}$
&\nodata& pow &B\\

LMC X-1         & 84.8917 & -69.7210 & HXB     & 2.3$^{+0.22}_{-0.20}$ & \nodata & pow &B\\

PSR B0540-69.3 & 84.9878  & -69.3230 & Pulsar  & 1.85$^{+0.28}_{-0.26}$ 
&\nodata& pow &B\\

PKS 0537-286   & 84.9953  & -28.7029 & BLAZAR  &
$1.35^{+0.06}_{-0.08}$ &$<0.01$  &wabs*pow &B, A \\

PKS 0548-322    & 87.7165 & -32.2610  & BL Lac  
&  $ 1.8^{+0.032}_{-0.031}$ &$ 0.02^{+0.006}_{-0.005}$   & wabs*pow & B, X\\

NGC 2110      & 88.0411 & -7.4554 &  Sy2  & 
$1.62^{+0.01}_{-0.01}$/$0.47^{+0.02}_{-0.02}$  &
$4.0^{+0.13}_{-0.07}$ & wabs*(pow+ga) + bb & B, A, X\\

 LEDA 75476  & 89.5237  & -38.3799 & Sy1 & $1.74^{+0.017}_{-0.025}$/
$0.25^{+0.08}_{-0.05}$ &
$2.5^{+0.11}_{-0.17} $& wabs*(pow+ga)+bb &  B, A\\

ESO 490- G 26   & 100.0031 & -25.8931 & Sy1.2  &
 1.90$^{+0.05}_{-0.04}$ & 0.27$^{+0.005}_{-0.005}$ & wabs*pow & B, X \\

SWIFT  J0727.5-2406 & 111.8951 & -24.1039 &\nodata &1.53$^{+0.55}_{-0.54}$
&\nodata & pow & B\\

V* 441 Pup      & 112.1626 & -26.0696 & CV & 
$12.4^{+13.6}_{-5.6}$ & \nodata & bremss & B \\

V* BG CMi      & 112.8752 & 9.9214  & CV  &  31.3$^{+41.2}_{-14.2}$& \nodata & bremss& B \\

SWIFT J0732.5-1331 & 113.1328 & -13.5037 & CV  & $33.2^{+50.1}_{-14.2}$ 
& \nodata & bremss & B \\

SWIFT J0739.6-3144 & 114.9127 & -31.7496 & Sy2\tablenotemark{d}   &$1.77^{+0.51}_{-0.43}$ 
& $>$2\tablenotemark{c} & pow & B\\

SWIFT J0743.0-2543 & 115.7501 & -25.7314 & \nodata & $1.78^{+0.69}_{-0.56}$ & 
\nodata & pow & B \\

IGR J07597-3842 & 119.9822 & -38.7422 & Sy1.2 & 
$1.8^{+0.08}_{-0.07}$& $<$0.01 &  wabs*pow &B, X\\

UGC 4203       & 121.0552 & 5.1203 & Sy2 &
1.68$^{+0.09}_{-0.10}$/ $0.31^{+0.08}_{-0.05}$ 
&12.5$^{+5.0}_{-3.7}$\tablenotemark{f} & wabs(pexrav+ga)+bb
 &B, A, X \\

SWIFT J0811.5+0937 &  122.8750 &  9.6214  &  XBONG\tablenotemark{d} &
$2.2^{+2.1}_{-0.9}$ & 
{$ ~0$}\tablenotemark{e} & pow &  B \\

SWIFT J0823.4-0457 & 125.8271 &  -4.9401 &  Sy2 &
$1.84^{+0.28}_{-0.22}$ 
&$19.3^{+6.8}_{-5.4}$ & wabs*pow & B, X\\

Vela PSR   & 128.8308 & -45.1771  & PSR & 1.88$^{+0.20}_{-0.26}$ & \nodata
& pow &B\\

FRL 1146        & 129.6151 & -35.9976  & Sy1  
& $1.88^{+0.37}_{-0.31}$ & \nodata & pow & B \\

3C 206 & 129.9556 &  -12.2467 &  QSO  &
$1.95^{+0.43}_{-0.39}$ & $0$\tablenotemark{e} 
& wabs*pow &  B \\

SWFIT  J0844.9-3531 & 131.2411 & -35.5313  &\nodata& $1.91^{+0.46}_{-0.68}$
 &\nodata & pow & B \\

SWIFT J0854.7+1502 & 133.6828 &  15.0371 & Sy2\tablenotemark{d}
  &$1.41^{+0.7}_{-0.9}$ &$  >0.5$\tablenotemark{c} & pow & B\\

SWIFT J0917.2-6221     & 139.112 & -62.359 & Sy1 & $1.87^{+0.07}_{-0.04}$/ 0.14$^{+0.02}_{-0.02}$
 & 1.33$^{+0.18}_{-0.10}$ & bb+wabs*pow & B \\

Mrk 0704  &  139.6505 & 16.2987 & Sy1.5
&$ 1.36^{+0.1}_{-0.07}$ & $15.0^{+6.3}_{-3.5}$ & pcfabs*(pow+ga) & B, A, X \\

4U 0919-54      & 140.0753 & -55.2135 & LXB & 45.11$^{+26.13}_{-9.8}$
 & \nodata & bremss &B\\

MCG -01-24-012 & 140.2134 & -8.0872 &  Sy2 &
$ 1.7^{+0.08}_{-0.07}$  & $6.5^{+0.8}_{-0.7}$ & wabs*pow & B, X\\

NGC 2992  & 146.4060 & -14.3007  & Sy1.9
 &$1.24^{+0.06}_{-0.05}$ 
&$0.17^{-0.03}_{-0.03}$ & wabs*(pow+ga)& B, A, X\\

ESO 434- G 040  & 146.9151 &-30.9388 &  Sy2  &
$1.77^{+0.006}_{-0.07}$/ $0.15^{+0.011}_{-0.016}$
 &$1.5^{+0.026}_{-0.09}$ &wabs*(pow+ga)+bb & B, A, X\\

3C 227   & 146.9447 & 7.4191 & Sy1 &
$2.11^{+0.14}_{-0.24}$ & $ 3.6$ & pexrav+wa*pow & B, C \\

NGC 3081  &  149.8805 &  -22.8561  &  Sy2   &
$1.9^{+0.02}_{-0.04}$/ $0.58^{+0.15}_{-0.11}$ 
&$60^{+3.1}_{-3.1}$ &wabs*(pow+ga) +bb  & B, A, S \\

\enddata
\tablenotetext{a}{Photon index and/or plasma temperature for the model,
specified in column ``Model'', to fit the data.}

\tablenotetext{b}{Instrumenst used for spectral analysis are: 
B = BAT, X = Swift/XRT, A = ASCA, C = Chandra, and S = BeppoSAX.}

\tablenotetext{c}{\ Lower limit on absorption estimated through 
the non detection by ROSAT}.

\tablenotetext{d}{Proposed identification in \cite{rau07}.}

\tablenotetext{e}{Order of magnitude of the absorption estimated imposing
that the extrapolated source flux match the ROSAT-PSPC count rates.}

\tablenotetext{f}{UGC 4203 exhibits transition between reflection-dominated
and a trasmission-dominated spectrum. The absorption is estimated in the
latter case using XRT data (see text for details).}

\tablerefs{References: (1) Gallo et al. (2005). }
\end{deluxetable}

%


\begin{deluxetable}{lcc}
\tablewidth{0pt}
\tablecaption{Spectral parameters for Sy1, Sy2, intermediate and  all Seyfert
AGN. Errors are 90\% confidence level.
\label{tab:class}}
\tablehead{
\colhead{CLASS}   & \colhead{Photon index}    &\colhead{$\chi^2$}/NDF\\
\colhead{} & \colhead{}  &\colhead{}
}
\startdata 
Seyfert 1       & 2.23$\pm0.11$  & 5.4/4\\
Seyfert 2       & 1.86$\pm0.10$  & 1.2/4 \\
Seyfert 1.2-1.5 & 1.95$\pm 0.11$  &4.9/4\\
Seyfert All     & 2.00$\pm 0.07$ & 2.1/4\\
\enddata
\end{deluxetable}
%
\begin{deluxetable}{lccccccccc}
\rotate
\tabletypesize{\scriptsize}
\tablewidth{0pt}
\tablecaption{Extragalactic sample. \label{tab:sample}}
\tablehead{
\colhead{NAME}         & \colhead{Type}   &\colhead{z}     & 
\colhead{Radio-loudness}   &
\colhead{Fx}           &
\colhead{Lx}           & 
\colhead{L$_{\textup {2-10\,keV}}$/L$_{\textup {OIII}}$\tablenotemark{a}}& 
 \colhead{Fe$_{EW}$\tablenotemark{b}} & \colhead{N$_H$}  &\colhead{ref}ß  \\
\colhead{} & \colhead{}  & \colhead{}    &\colhead{}  &
\colhead{\scriptsize ($10^{-11}$\,erg cm$^{-2}$ s$^{-1}$)}&
\colhead{\scriptsize ($10^{43}$\,erg s$^{-1}$) } & \colhead{} &
\colhead{(eV)} &
\colhead{\scriptsize ($10^{22}$atoms cm$^{-2}$) } & \colhead{}
} 
\startdata 
%

3C 105.0           & Sy2    & 0.089   & 28421  
& 4.6$^{+0.5}_{-0.4}$ &  44.5$^{+17}_{-22}$ & 581.5$\pm69.2$
&     nr &  29.4 & 1\\

1AXG J042556-5711  & Sy1    & 0.104   & 0.78\tablenotemark{c} &  
1.92$^{+0.3}_{-0.2}$ & 55.0$^{+10}_{-8}$ &
\nodata
  & nr & 0 & 1\\

3C 120             & Sy1    & 0.0330  & 3762  
&10.1$^{+0.8}_{-1.8}$   & 25.4$^{+2.2}_{+4.5}$  & 148.8$\pm40.6$ &
 52.3 & 0   & 1\\

MCG -01-13-025     & Sy1.2  &0.015894 & 1.15\tablenotemark{c} & 
2.52$^{+1.1}_{-1.6}$  & 1.5$^{+0.6}_{-0.8}$ &  \nodata&
\nodata & $<$0.02 & 2\\

SWIFT J0505.7-2348 & Sy2    & 0.0350  & 7.13\tablenotemark{c} &  
5.0$^{+0.7}_{-1.5}$ &  14.1$^{+0.24}_{-5.0}$  & 345.2$\pm82.2$
& nr   & 6.3 & 1\\

QSO B0513-002      & Sy1    & 0.0327  & 0.254  & 
4.92$^{+0.9}_{-2.8}$  & 12.3$^{+2.1}_{-7.3}$  & \nodata
&90.8 &  0.02 & 3\\

ESO 362- G 018     & Sy1.5  & 0.0126  & 0.58\tablenotemark{c} & 
5.0$^{+0.7}_{-1.0}$  & 1.7$^{+0.3}_{-0.3}$  &  36.3$\pm3.7$
&  nr &  $<$0.01   & 1 \\

Pictor A           & Sy1    & 0.035   & 14045  & 
1.8$^{+0.4}_{-1.2}$  & 5.1$^{+1.5}_{-3.7}$  & 113.4$\pm14.5$
 & nr & 0.12     &  1  \\

ESO 362-G021       & BL Lac & 0.05534 & 2409   
& 2.7$^{+0.3}_{-0.3}$  &  18.8$^{+5.0}_{-0.5}$ & 1284.1$\pm270.3$
&nr  & 0.1     & 1 \\

PKS 0537-286       & BLAZAR & 3.1     & 22000  
& 2.43$^{+0.9}_{-3.0}$  & 1.2$^{+0.4}_{-0.4}\times 10^5$ & \nodata
 & nr & $<$0.01 & 1 \\

PKS 0548-322       & BL Lac & 0.0690  & 383.33  
&3.1$^{+0.6}_{-1.0}$   & 37.0$^{+6.0}_{-11.0}$ & \nodata
 & nr   & 0.0257     & 1 \\

NGC 2110           & Sy2    & 0.007789& 26.92  
&27.0$^{+0.9}_{-1.0}$ & 3.5$^{+0.1}_{-0.1}$ & 1115.8$\pm101.4$
& 118  & 4.0 & 1 \\

LEDA 75476         & Sy1    & 0.0338  & 2.87\tablenotemark{c} 
& 3.2$^{+0.6}_{-1.2}$  &  8.7$^{+2.4}_{-2.1}$ & 393.7$\pm59.6$
 &144   & 2.5   & 1 \\

UGC 4203           & Sy2    & 0.01349 & 7.67  & 
4.28$^{+0.6}_{-1.4}$ & 1.7$^{+0.3}_{-0.6}$   & 214.2$\pm71.4$
& 747 & 12.5 & 1 \\

SWIFT J0811.5+0937 & XBONG  & 0.282   & 268\tablenotemark{c} & 
1.55$^{+1.2}_{-1.54}$ & 384$^{+170}_{-200}$ & \nodata  &\nodata
 & $\sim0$\tablenotemark{d} & \nodata\\

SWIFT J0823.4-0457 & Sy2    & 0.023   & 0.61\tablenotemark{c}& 
2.78$^{+1.0}_{-1.1}$ & 3.3$^{+1.1}_{-1.3}$ & 179.1$\pm71.6$ &
\nodata   & 16.2 & 1\\

3C 206             & QSO    & 0.1976  & 1194  & 
2.62$^{+0.7}_{-1.3}$ & 300$^{+75}_{-116}$ &\nodata 
& \nodata &0  & 1 \\

SWIFT J0854.7+1502 & Sy2    & 0.0696  & \nodata  & 
1.73$^{+1.2}_{-1.5}$ & 19.7$^{+13}_{-16}$  &\nodata 
& \nodata   & $>0.5$\tablenotemark{d} &1 \\

Mrk 0704           & Sy1    & 0.0292  & 0.82\tablenotemark{c} 
& 2.21$^{+1.1}_{-0.9}$   & 4.3$^{+2.1}_{-1.8}$ & 114.9$\pm27.1$
& 160& $14.6$ & 1\\

MCG -01-24-012     & Sy2    & 0.01964 & 2.86\tablenotemark{c} & 
4.6$^{+0.7}_{-0.9}$ & 3.7$^{+0.4}_{-0.6}$ &2914.4$\pm1092.9$
  & nr & 6.8  & 1 \\

NGC 2992           & Sy1.9  & 0.00771 & 2.03 &
3.6$^{+1.0}_{-1.1}$   & 0.47$^{+0.13}_{-0.10}$& 0.8$\pm0.1$
 & 520 & 0.17 & 1 \\

ESO 434- G 040     & Sy2    & 0.00848 & 0.6  & 
19.1$^{+0.6}_{-0.6}$ & 2.7$^{+0.1}_{-0.1}$ &  912.9$\pm21.1$
&85.5 & 1.5      & 1 \\

3C 227             & Sy1    & 0.0858  & 5462  &
2.23$^{+0.3}_{-1.6}$ &  40.0$^{+1.0}_{-3.2}$  & 19.3$\pm7.10$ &nr
 & 3.6&   1 \\

NGC 3081           & Sy2    & 0.00798  & 0.1\tablenotemark{c} & 
6.8$^{+0.9}_{-0.8}$  & 0.96$^{+0.1}_{-0.11}$ & 31.0$\pm1.7$
  &241    & 60 & 1  \\

\enddata
\tablenotetext{a}{The OIII luminosities have been derived in \cite{rau07}.}
\tablenotetext{b}{Iron line equivalent width. A value of ``nr'' means
that the iron line is statistically not required by the fit.}
\tablenotetext{c}{The radio flux at other wavelengths 
has been extrapolated to 6 cm
assuming $f_{\nu} \propto {\nu}^{-0.5}$.}
\tablenotetext{d}{\ Limit on the absorption obtained estrapolating the BAT
spectrum to the ROSAT band.
}
\tablerefs{References for the absorption values: 
(1) this work;  (2) Gallo et al. (2005);  (3) Lutz et al. (2004).
}
\end{deluxetable}
\begin{deluxetable}{lcccc}
\tablewidth{0pt}
\tablecaption{Comparison with previous results\label{tab:logn_comp}}
\tablehead{
\colhead{Instrument} & \colhead{Ref.}   & \colhead{Energy}    
& \colhead{AGN density\tablenotemark{a}} &\colhead{BAT density (this work)\tablenotemark{b}} \\
\colhead{} & & \colhead{keV} & \colhead{10$^{-2}$ deg$^{-2}$}&
\colhead{10$^{-2}$ deg$^{-2}$} 
}
\startdata 

INTEGRAL-ISGRI  & 1 & 20  - 40  &  0.48$\pm0.08$    & 0.41$\pm 0.08$\\
INTEGRAL-ISGRI  & 2 & 100 - 150 & 0.18$\pm0.006$    & 0.17$\pm 0.034$\\    
HEAO-1 A2       & 3 & 2   - 10  & 1.2$\pm0.2$       & 1.6 $\pm0.32$ \\
RXTE PCA        & 4 & 8   - 20  & 0.56$\pm0.06$     & 0.65$\pm0.13$\\   
XMM             & 5 & 0.5 - 2   & 0.1$\pm0.01$      & 1.3 $\pm0.26$\\
XMM             & 5&  2   - 10  & 0.95$\pm0.06$     & 1.6 $\pm0.32$ \\
XMM             & 5&  5   - 10  & 0.63$\pm0.4$      & 0.4$\pm0.08$\\   
\enddata
\tablenotetext{a}{AGN densities from different surveys above 10$^{-11}$ 
erg cm$^{-2}$ s$^{-1}$ (in the respective bands).}
\tablenotetext{b}{The BAT AGN density was converted to the native energy
band of the measurement we are comparing it with.}
\tablerefs{(1)Beckmann et al. 2006; (2) Bazzano et al. 2006; 
(3) Piccinotti et al. 1982; (4) Revnivtsev et al. 2004; 
(5) Cappelluti et al. 2007}
\end{deluxetable}


\clearpage


\begin{figure*}
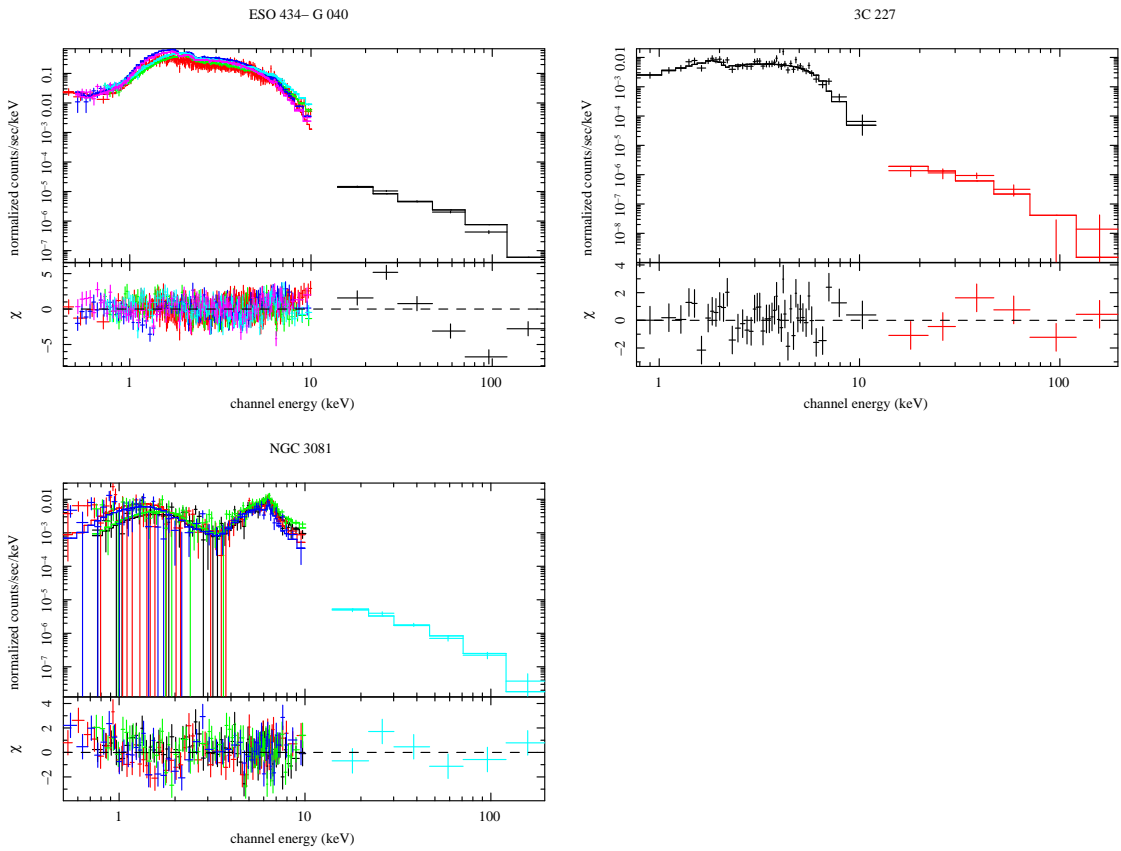
 
    \begin{center}
     \begin{tabular}{cc}
  \includegraphics[scale=0.3,angle=270]{f14a.ps} &
    \includegraphics[scale=0.3,angle=270]{f14b.ps}\\
\includegraphics[scale=0.3,angle=270]{f14c.ps} &
     \end{tabular}
    \end{center}
    \null\vspace{-7mm}
    \caption{{\bf (for online version)}
Folded spectra and best fit models as described in the text. 
From left to right and up to bottom the spectra are for: 
ESO 434 -G 040, 3C 227 
and NGC 3081.
}
  \label{fig:spe8}
\end{figure*}

\end{document}